\documentclass[aps,prd,twocolumn,superscriptaddress]{revtex4}

\usepackage{epsfig}
\usepackage{enumitem}

\usepackage{graphics}     
\usepackage{graphicx}     
\usepackage{subfigure}
\usepackage{longtable}     
\usepackage{url}          
\usepackage{bm}           
\usepackage{natbib}
\usepackage{textpos}
\usepackage{amsfonts,amsmath,amssymb,mathrsfs,bm,amsthm}
\usepackage{float}
\usepackage{booktabs}
\usepackage{array}
\usepackage{tabu}
\usepackage{dcolumn}
\usepackage{rotating}
\usepackage{ulem}

\newcommand{\RN}[1]{%
  \textup{\uppercase\expandafter{\romannumeral#1}}%
}

\usepackage{nicefrac}
\usepackage{amsthm}
\usepackage{color}
\usepackage{cancel}
\usepackage{appendix}
\usepackage{makecell}
\usepackage{upgreek}

\usepackage[colorlinks=true,linkcolor=black,citecolor=blue,urlcolor=blue,bookmarksopen]{hyperref}

\newcommand{\appropto}{\mathrel{\vcenter{
  \offinterlineskip\halign{\hfil$##$\cr
    \propto\cr\noalign{\kern2pt}\sim\cr\noalign{\kern-2pt}}}}}
\renewcommand{\v}[1]{\boldsymbol{#1}}		

\begin{document}

\title{Novel signatures of dark matter in laser-interferometric gravitational-wave detectors}

\author{H.~Grote}
\email{groteh@cardiff.ac.uk}
\affiliation{Cardiff University, School of Physics and Astronomy, The Parade, CF24 3AA, United Kingdom}

\author{Y.~V.~Stadnik}
\email{yevgenystadnik@gmail.com}
\affiliation{Helmholtz Institute Mainz, Johannes Gutenberg University, 55128 Mainz, Germany}
\affiliation{Kavli Institute for the Physics and Mathematics of the Universe (WPI), The University of Tokyo Institutes for Advanced Study, The University of Tokyo, Kashiwa, Chiba 277-8583, Japan}

\raggedbottom

\date{\today}

\begin{abstract}
Dark matter may induce apparent temporal variations in the physical ``constants'', including the electromagnetic fine-structure constant and fermion masses. 
In particular, a coherently oscillating classical dark-matter field may induce apparent oscillations of physical constants in time, while the passage of macroscopic dark-matter objects (such as topological defects) may induce apparent transient variations in the physical constants. 
In this paper, we point out several new signatures of the aforementioned types of dark matter that can arise due to the geometric asymmetry created by the beam-splitter in a two-arm laser interferometer. 
These new signatures include dark-matter-induced time-varying size changes of a freely-suspended beam-splitter and associated time-varying shifts of the main reflecting surface of the beam-splitter that splits and recombines the laser beam, as well as time-varying refractive-index changes in the freely-suspended beam-splitter and time-varying size changes of freely-suspended arm mirrors. 
We demonstrate that existing ground-based experiments already have sufficient sensitivity to probe extensive regions of unconstrained parameter space in models involving oscillating scalar dark-matter fields and domain walls composed of scalar fields. 
In the case of oscillating dark-matter fields, Michelson interferometers --- in particular, the GEO\,600 detector --- are especially sensitive. 
The sensitivity of Fabry-Perot-Michelson interferometers, including LIGO, VIRGO and KAGRA, to oscillating dark-matter fields can be significantly increased by making the thicknesses of the freely-suspended Fabry-Perot arm mirrors different in the two arms. 
Not-too-distantly-separated laser interferometers can benefit from cross-correlation measurements in searches for effects of spatially coherent dark-matter fields. 
In addition to broadband searches for oscillating dark-matter fields, we also discuss how small-scale Michelson interferometers, such as the Fermilab holometer, could be used to perform resonant narrowband searches for oscillating dark-matter fields with enhanced sensitivity to dark matter. 
Finally, we discuss the possibility of using future space-based detectors, such as LISA, to search for dark matter via time-varying size changes of and time-varying forces exerted on freely-floating test masses.


\end{abstract}


\maketitle

\section{Introduction}
\label{Sec:Intro}

While the existence of dark matter (DM) is well established from astrophysical and cosmological observations, the elucidation of its precise nature remains one of the most important problems in contemporary physics. 
Since extensive searches for DM particles of relatively high masses (e.g., WIMPs) through their possible non-gravitational effects have not yet produced a strong positive result, in recent years the possibility of searching for low-mass (sub-eV) DM candidates has been receiving increased attention. 
There are numerous well-motivated DM candidates of this type, including the canonical axion, axion-like particles and dilatons, which may form a coherently oscillating classical field and/or stable solitonic field configurations known as topological defects (such as domain walls). 
Searches for such low-mass DM candidates via possible particlelike signatures (such as recoils, energy depositions and ionisations) are practically impossible, since the individual non-relativistic DM particles in this case carry very small momenta. 
Instead, one can take advantage of the fact that low-mass DM particles must have large occupation numbers if they comprise the observed DM content of the Universe (the average local cold DM density is given by $\rho_\textrm{DM} \approx 0.4~\textrm{GeV}/\textrm{cm}^3$ \cite{PDG_2018_review}) and look for wavelike and other coherent signatures of these DM fields. 
In recent years, a number of novel ideas have emerged to search for low-mass DM using precision measurement techniques from the fields of atomic and optical physics; see \cite{Stadnik_2018_review,Safronova_2018_review} for recent overviews. 

DM may induce apparent temporal variations in the physical ``constants'', including the electromagnetic fine-structure constant $\alpha$, as well as the electron and nucleon masses $m_e$ and $m_N$, via certain non-gravitational interactions with standard-model (SM) fields \cite{Footnote1}. 
In particular, a coherently oscillating classical DM field may induce apparent oscillations of physical constants in time \cite{Stadnik_2015_DM-LI,Stadnik_2015_DM-VFCs}, while the passage of topological defects may induce apparent transient variations in the physical constants \cite{Derevianko_2014_TDM,Stadnik_2014_TDM}. 
Some possible effects of such time-varying physical constants in laser interferometers and optical cavities, including time-varying changes of solid sizes and laser frequencies, were explored in \cite{Stadnik_2015_DM-LI,Stadnik_2016_cavities}. 
Several clock-cavity comparison experiments searching for DM-induced time-varying physical constants have been conducted recently \cite{Wcislo_2016_cavity-DM,Ye_2018_cavity-DM,Aharony_2019_cavity-DM}. 
Experiments of this type are mainly sensitive to oscillation frequencies up to $\sim\textrm{Hz}$ (equivalently timescales down to $\sim\textrm{s}$). 
However, we would also like to precisely probe even higher oscillation frequencies (audio-band frequencies and beyond), which too are interesting from the point of view of current astrophysical observations. 

\begin{figure}[h!]
\centering
\includegraphics[width=1.0\linewidth]{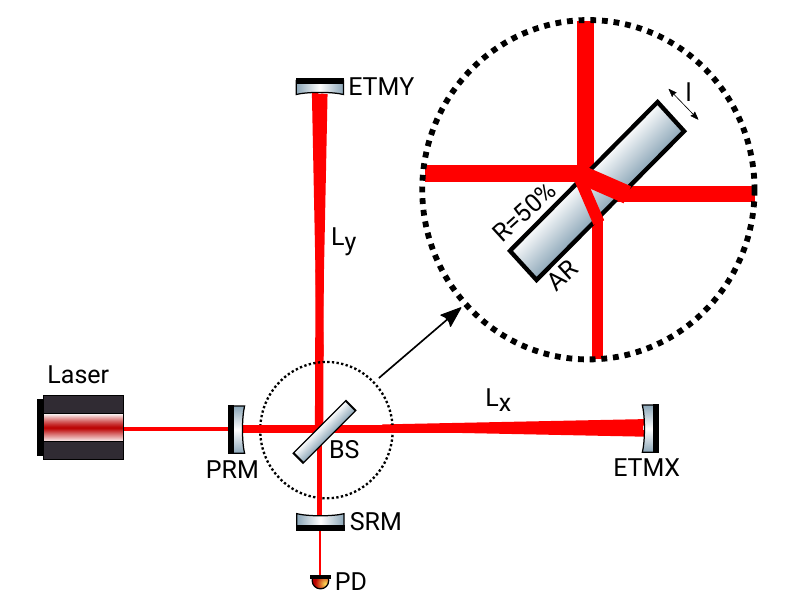}
\caption{ 
Simplified layout of a dual-recycled Michelson interferometer, such as GEO\,600 \cite{GEO600_2013,GEO600_2015} and the Fermilab holometer \cite{Fermilab_holometer_2016,Fermilab_holometer_2017}. 
Dual recycling denotes the combination of power recycling and signal recycling. 
PRM:~power recycling mirror;
BS:~beam-splitter of thickness $l$; 
ETMX, ETMY:~end arm mirrors (test masses); 
SRM:~signal recycling mirror; 
PD:~photodetector. 
The inset shows the beam routing through the beam-splitter. 
The beam-splitting surface typically has a power reflectivity of $R = 50\%$. 
The opposing face of the beam-splitter, denoted by AR, is anti-reflective coated. 
For clarity, we have omitted the single folding of the arms in GEO\,600, as well as the second co-located interferometer of the Fermilab holometer; furthermore, the Fermilab holometer does not have a signal recycling mirror SRM. 
}
\label{Fig:GEO_config}
\end{figure}

\begin{figure}[h!]
\centering
\includegraphics[width=1.0\linewidth]{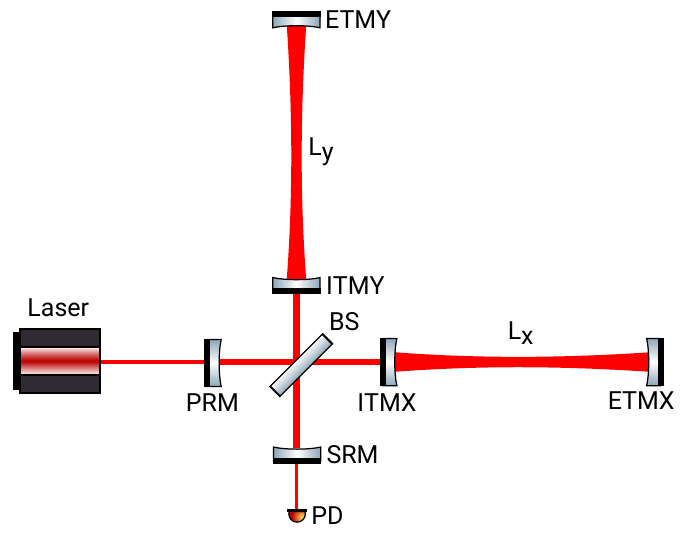}
\caption{
Simplified layout of a dual-recycled Fabry-Perot-Michelson interferometer, such as Advanced LIGO \cite{aLIGO_2016}, VIRGO \cite{aVirgo_2014} and KAGRA \cite{KAGRA_2013}. 
PRM:~power recycling mirror; 
BS:~beam-splitter; 
ITMX, ETMX, ITMY, ETMY:~arm mirrors (test masses); 
SRM:~signal recycling mirror; 
PD:~photodetector. 
The arm mirrors (test masses) are separated by the distances $L_x$ and $L_y$, which are 4 km in the case of LIGO.} 
\label{Fig:LIGO_config}
\end{figure}

In this work, we propose new ways of searching for DM with laser interferometers. 
A two-arm laser interferometer is typically used to detect small changes in the difference of the optical path lengths in the two arms of the interferometer. 
Since the two arms of an interferometer are practically equal in terms of optical path length, time-varying arm length fluctuations and changes in the laser frequency due to a homogeneous DM field are common mode, and their effects are therefore strongly suppressed in the output signal. 
There is, however, a geometric asymmetry created by the beam-splitter in a two-arm laser interferometer. 
We point out that the beam-splitter and arm mirrors of an interferometer, if freely suspended, can produce differential optical-path-length changes if one or more of the physical constants of nature vary in time (and space). 
A non-zero output signal, namely a phase difference between the two arms of the interferometer, can arise in several ways. 
If the DM field is homogeneous across the entire interferometer, then the main observable effect will generally arise from the freely-suspended beam-splitter. 
A freely-suspended beam-splitter would experience time-vaying size changes about its centre-of-mass, thus shifting back-and-forth the main reflecting surface that splits and recombines the laser beam (see the inset in Fig.~\ref{Fig:GEO_config}). 
Additionally, (generally smaller) time-varying changes in the refractive index of the beam-splitter would change the optical path length across the beam-splitter. 
On the other hand, if the DM field is inhomogeneous over an interferometer, then substantial observable effects may also arise from time-varying size changes of the freely-suspended arm mirrors (see Figs.~\ref{Fig:GEO_config} and \ref{Fig:LIGO_config}). 
In some situations, the output signal can be significantly enhanced if the arm mirrors have different physical characteristics (in particular, different thicknesses).

Laser interferometry has been optimised over decades to develop ultra-sensitive gravitational-wave detectors, which have recently been employed spectacularly to directly observe gravitational waves on Earth for the first time \cite{LIGO_discovery_2016,LIGO_discovery_2017}. 
Additionally, smaller-scale interferometers have more recently been utilised to search for non-gravitational-wave phenomena, such as quantum-geometry effects that may arise at the Planck scale \cite{Fermilab_holometer_2016,Fermilab_holometer_2017}. 
Future space-based laser-interferometric gravitational-wave detectors, such as LISA \cite{LISA_2017}, are currently under development. 
In this paper, we explore novel signatures of DM in ground- and space-based laser interferometers. 
We estimate the sensitivities of these detectors to the physical parameters of models of DM consisting of a coherently oscillating classical field or domain walls. 
Searches for coherently oscillating classical DM fields share similarities with searches for continuous as well as for stochastic gravitational waves, while searches for domain-wall objects share similarities with searches for gravitational-wave bursts. 
Based on our estimates, we emphasise that existing laser interferometers, particularly the GEO\,600 Michelson interferometer \cite{GEO600_2013,GEO600_2015}, already have sufficient sensitivity to probe extensive regions of unconstrained parameter space in these models, including scalar DM fields oscillating at frequencies in the range $\sim100~\textrm{Hz} - 10~\textrm{kHz}$ and scalar-field domain walls with transverse sizes of up to several km. 
The sensitivity of Fabry-Perot-Michelson interferometers, including LIGO \cite{aLIGO_2016}, VIRGO \cite{aVirgo_2014} and KAGRA \cite{KAGRA_2013}, to oscillating DM fields can be significantly increased by making the thicknesses of the freely-suspended Fabry-Perot arm mirrors different in the two arms. 
In this case, the sensitivity of these experiments to conventional gravitational-wave searches, which can be performed simultaneously with our suggested DM searches, would not necessarily be degraded. 
Not-too-distantly-separated laser interferometers can benefit from cross-correlation measurements in searches for effects of spatially coherent dark-matter fields. 
In addition to broadband searches for oscillating DM fields, we also discuss how small-scale Michelson interferometers could be used to perform resonant narrowband searches for oscillating DM fields with enhanced sensitivity to underlying DM interactions. 

The structure of our paper is as follows. 
In Sec.~\ref{Sec:Theory}, we discuss how DM can induce apparent temporal variations in the physical constants and derive the effects of such variations on freely-suspended beam-splitters, as well as freely-suspended and freely-floating test masses. 
In Sec.~\ref{Sec:ODM}, we consider the specific model of a coherently oscillating classical DM field; we derive the effects of an oscillating DM field on ground- and space-based laser interferometers and estimate the sensitivities of existing, modified and future experiments to the underlying DM parameters. 
In Sec.~\ref{Sec:TDM}, we consider the specific model of topological defects in the form of domain walls; we derive the effects of domain walls on ground- and space-based laser interferometers and estimate the sensitivities of existing and future experiments to the underlying DM parameters. 
Finally, in Sec.~\ref{Sec:Discussion}, we summarise our findings and discuss DM searches with laser interferometers in the context of other measurements. 

Throughout this work, unless explicitly stated otherwise, we shall adopt the natural system of units $\hbar = c =1$, where $\hbar$ is the reduced Planck constant and $c$ is the speed of light in vacuum. 
In this paper, we express the interferometer output in terms of the difference of the optical path lengths in different arms of an interferometer. 

\section{Theory and effects of dark-matter-induced varying physical ``constants''}
\label{Sec:Theory}

\subsection{Non-gravitational interactions of scalar fields}
\label{Sec:DM_interactions}
A scalar (spinless, even-parity) field $\phi$ can couple to the SM fields in a number of possible ways. 
Generally, the simplest possibility involves linear-in-$\phi$ interactions: 
\begin{equation}
\label{lin_portal}
\mathcal{L}_\textrm{int}^\textrm{lin} = \frac{\phi}{\Lambda_\gamma} \frac{F_{\mu\nu}F^{\mu\nu}}{4} - \sum_f \frac{\phi}{\Lambda_f} m_f \bar{f}f \, ,
\end{equation}
where the first term represents the coupling of the scalar field to the electromagnetic field tensor $F$, while the second term represents the coupling of the scalar field to the SM fermion fields $f$, with $m_f$ the ``standard'' mass of the fermion and $\bar{f} = f^\dagger \gamma^0$ the Dirac adjoint. 
The linear couplings in (\ref{lin_portal}) can be generated, e.g., via the super-renormalisable interaction of $\phi$ with the Higgs field; see \cite{Piazza_2010_Higgs,Stadnik_2016_Higgs} for more details. 
These linear couplings, however, may be absent, e.g., as a result of an underlying $Z_2$ symmetry (invariance under the transformation $\phi \to -\phi$). 
In this case, the simplest possibility would involve quadratic-in-$\phi$ interactions: 
\begin{equation}
\label{quad_portal}
\mathcal{L}_\textrm{int}^\textrm{quad} = \left(\frac{\phi}{\Lambda'_\gamma}\right)^2 \frac{F_{\mu\nu}F^{\mu\nu}}{4} - \sum_f \left(\frac{\phi}{\Lambda'_f}\right)^2 m_f \bar{f}f \, .
\end{equation}

Comparing the terms in Eqs.~(\ref{lin_portal}) and (\ref{quad_portal}) with the relevant terms in the SM Lagrangian: 
\begin{equation}
\label{SM_Lagrangian}
\mathcal{L}_\textrm{SM} \supset - \frac{F_{\mu\nu}F^{\mu\nu}}{4} - \sum_f  q_f J_\mu A^\mu - \sum_f  m_f \bar{f}f \, ,
\end{equation}
where $q_f$ is the electric charge carrried by the fermion $f$, $J^\mu = \bar{f} \gamma^\mu f$ is the electromagnetic 4-current and $A^\mu$ is the electromagnetic 4-potential, 
we see that the linear interactions in (\ref{lin_portal}) effectively alter the fine-structure constant and fermion masses according to: 
\begin{equation}
\label{lin_VFCs}
\alpha \to \frac{\alpha}{1 - \phi/\Lambda_\gamma} \approx \alpha \left( 1 + \frac{\phi}{\Lambda_\gamma} \right) , \, ~ m_f \to m_f \left(1 + \frac{\phi}{\Lambda_f} \right) , \,
\end{equation}
while the quadratic interactions in (\ref{quad_portal}) effectively alter the constants according to: 
\begin{align}
\label{quad_VFCs}
\alpha \to \frac{\alpha}{1 - \left(\phi/\Lambda'_\gamma\right)^2} \approx \alpha \left[ 1 + \left(\frac{\phi}{\Lambda'_\gamma}\right)^2 \right] , \, \notag \\
m_f \to m_f \left[1 + \left(\frac{\phi}{\Lambda'_f}\right)^2 \right] . \,
\end{align}

\subsection{Size changes of beam-splitter and test masses}
\label{Sec:Lengths}
Time-varying $\alpha$ and particle masses alter the geometric sizes of solid objects. 
The length of a solid is given by $L \sim N a_\textrm{B}$, where $N$ is the number of lattice spacings and $a_\textrm{B} = 1/(m_e \alpha)$ is the atomic Bohr radius. 
In the adiabatic limit, the size of a solid body thus changes according to: 
\begin{equation}
\label{size_changes1}
\left(\frac{\delta L}{L}\right)_0 \approx \frac{\delta a_\textrm{B}}{a_\textrm{B}} = - \frac{\delta \alpha}{\alpha} - \frac{\delta m_e}{m_e} \, . 
\end{equation}
Additionally, there are also small relativistic corrections associated with electromagnetic processes and finite-nuclear-mass effects \cite{Stadnik_2015_DM-LI,Flambaum_2018_sizes}. 
The former affect the $\alpha$-dependence in Eq.~(\ref{size_changes1}) and scale roughly as $\propto \left(Z \alpha\right)^2$ with the nuclear charge $Z$, while the latter affect the $m_e/m_N$-dependence and are of the order $\sim m_e/m_N \approx 5 \times 10^{-4}$. 
We thus see that the relativistic corrections are negligibly small in most common materials (including silica and sapphire) and only become non-negligible in the heaviest stable elements [where their size is $\sim 20\%$ of the non-relativistic contribution in Eq.~(\ref{size_changes1})].

The expression in Eq.~(\ref{size_changes1}) is only valid in the adiabatic limit, when the solid can optimally respond to the slow perturbations induced by the DM field. 
To model more general perturbations, we can treat the response of a solid to perturbations associated with a particular Fourier-driving-frequency component $f$ within the simple model of a strongly-underdamped driven harmonic oscillator (damping parameter $\zeta \ll 1$), noting that the length perturbations of the solid are induced by an external field \cite{footnote_Sep2019}. 
In this case, the steady-state response of the solid size changes is given by: 
\begin{equation}
\label{steady-state_underdamped_solution}
\left(\frac{\delta L}{L}\right)_f = \left(\frac{\delta L}{L}\right)_0 \times \frac{1}{\sqrt{\left[1 - \left( f / f_0 \right)^2 \right]^2 + \left(2 \zeta f / f_0 \right)^2}} \, , 
\end{equation}
where $(\delta L / L)_0$ is the fractional size change in the adiabatic limit [see Eq.~(\ref{size_changes1})], and $f_0$ is the frequency of a fundamental vibrational mode of the solid. 
There are three limiting cases in Eq.~(\ref{steady-state_underdamped_solution}): 
\begin{itemize}
\item When $f \ll f_0$, the size changes of the solid are independent of the driving frequency and coincide with the adiabatic case in Eq.~(\ref{size_changes1}). 
The size oscillations of the solid are in phase with the oscillating DM field. 
\item When $f \approx f_0$, the size changes of the solid are enhanced by the large quality factor $Q_\textrm{mech} = 1/(2 \zeta) \gg 1$ compared with the adiabatic case. 
The size oscillations of the solid lag behind the oscillating DM field by a phase factor of $\pi/2$. 
\item When $f \gg f_0$, the size changes of the solid are suppressed by the factor $(f_0/f)^2 \ll 1$ compared with the adiabatic case. 
The size oscillations of the solid lag behind the oscillating DM field by a phase factor of $\pi$. 
\end{itemize}
All of the relevant optical components in ground-based gravitational-wave detectors are approximately cylindrically symmetric. 
In this case, the fundamental frequency of the longitudinal vibrational mode is given by: 
\begin{equation}
\label{fund_freq_long_mode}
f_0 = \frac{v_s}{2 L_i} \, ,
\end{equation}
where $v_s$ is the sound speed in the solid component (typically of the order of half-dozen~km/s in most commonly used materials) and $L_i$ is the length of the component.

\subsection{Refractive-index changes in beam-splitter}
\label{Sec:Beam-splitter}
Time-varying $\alpha$ and particle masses also affect the propagation of a light beam through the beam-splitter via alteration of the refractive index of the beam-splitter. 
To estimate the size of these effects, we assume that the laser (angular) frequency $\omega$ is much larger than all of the phonon-mode frequencies of the beam-splitter and adopt a simple Lorentz model with a single electronic mode of frequency $\omega_0$, in the regime of normal dispersion, $\omega_0 > \omega$. 
In this case, far away from the electronic resonance, the expression for the refractive index of a dielectric material reads:
\begin{equation}
\label{ri_definition}
n \approx \sqrt{\varepsilon_r} \, ,
\end{equation}
with the relative permittivity given by: 
\begin{align}
\label{Drude_permittivity}
\varepsilon_r &\approx 1 + \frac{\xi N \alpha}{m_e} \frac{1}{\omega_0^2 - \omega^2} \notag \\ 
&\approx 1 + \frac{\xi N \alpha}{m_e \omega_0^2} \left(1 + \frac{\omega^2}{\omega_0^2} \right) \, , 
\end{align}
where $N$ is the number density of atoms in the dielectric material and $\xi$ is a numerical constant that is independent of the physical constants. 
Let us first consider the non-dispersive term $\xi N \alpha / (m_e \omega_0^2)$ in Eq.~(\ref{Drude_permittivity}). 
Since $N \propto 1/a_\textrm{B}^3 = \left(m_e \alpha \right)^3$ and $\omega_0 \propto m_e \alpha^2$, the combination of parameters $\xi N \alpha / (m_e \omega_0^2)$ is independent of the physical constants. 
Hence the main effects of varying physical constants on the index of refraction arise through the dispersive term $\omega^2/\omega_0^2$ in (\ref{Drude_permittivity}): 
\begin{equation}
\label{ri_VFCs_general}
\frac{\delta n}{n} \approx \frac{\omega}{n} \frac{\partial n}{\partial \omega} \left( \frac{\delta \omega}{\omega} - \frac{\delta \omega_0}{\omega_0}  \right)  \approx \frac{\omega}{n} \frac{\partial n}{\partial \omega} \left( \frac{\delta \omega}{\omega} - 2 \frac{\delta \alpha}{\alpha} - \frac{\delta m_e}{m_e}  \right) \, . 
\end{equation}

Most experiments use a laser of wavelength $\lambda \approx 1~\mathrm{\upmu}\textrm{m}$ and a silica beam-splitter, for which $n \approx 1.5$ and $\omega/n \cdot \partial n / \partial \omega \approx 5 \times 10^{-3}$. 
If the laser is stabilised to a high-finesse reference cavity, in which the length of the solid spacer between the mirrors is allowed to vary naturally, and the cavity length changes are adiabatic, then by Eq.~(\ref{size_changes1}) we have $\delta \omega / \omega = - \delta L_\textrm{cav} / L_\textrm{cav} \approx \delta \alpha / \alpha + \delta m_e / m_e$, giving: 
\begin{equation}
\label{ri_VFCs_reference_cavity-optimal}
\frac{\delta n}{n} \approx -5 \times 10^{-3}  \frac{\delta \alpha}{\alpha}  \, . 
\end{equation}
On the other hand, if the laser is stabilised to a high-finesse reference cavity, in which the cavity length is independent of the length of the spacer between the mirrors (e.g., through the use of a multiple-pendulum suspension system for the mirrors), or if the cavity length changes are sufficiently rapid [see Eq.~(\ref{steady-state_underdamped_solution})], then $\delta \omega / \omega$ is approximately independent of changes in the physical constants. 
In this case, we instead have: 
\begin{equation}
\label{ri_VFCs_reference_cavity-suppressed}
\frac{\delta n}{n} \approx -5 \times 10^{-3} \left( 2 \frac{\delta \alpha}{\alpha} + \frac{\delta m_e}{m_e} \right) \, . 
\end{equation}
In all existing ground-based gravitational-wave detectors, the laser is ultimately stabilised to the common-mode interferometer arm length, which is isolated against length fluctuations (via the suspension points) via multiple-pendulum suspension systems for the mirrors, and so in this case Eq.~(\ref{ri_VFCs_reference_cavity-suppressed}) applies \cite{Footnote0}.

\subsection{Centre-of-mass displacements of test masses}
\label{Sec:CM_test_masses}
By analogy with the acceleration that a test particle or test mass experiences in the presence of a spatial gradient in a potential, spatial gradients in $\alpha$ and the particle masses give rise to accelerations on test particles and test masses of mass $M_\textrm{test}$ \cite{Footnote0B}: 
\begin{equation}
\label{test_mass_acceleration_general}
\delta \v{a}_\textrm{test} = - \frac{\v{\nabla} M_\textrm{test}}{M_\textrm{test}} \, . 
\end{equation}
The overall mass of an atom with $Z \gg 1$ consists of three different types of contributions:
\begin{equation}
\label{Bethe-Weizsaecker_formula}
M_\textrm{atom} \approx A m_N + Z m_e + \frac{a_C Z^2}{A^{1/3}} \, , 
\end{equation}
where $A$ is the total nucleon number of the nucleus. 
The last term in (\ref{Bethe-Weizsaecker_formula}) denotes the energy associated with the electrostatic repulsion between protons in a spherical nucleus of uniform electric-charge density, with $a_C \approx 0.7~\textrm{MeV}$. 

Most ground-based experiments employ beam-splitters and test masses made of silica. 
In this case, the relative contributions to the total test mass from the nucleon masses, electron mass and Coulomb energy are $\approx 1$, $\approx 3 \times 10^{-4}$ and $\approx 1.4 \times 10^{-3}$, respectively. 
LISA employs Au-Pt alloy ($\approx$~60:40 ratio) test masses. 
In this case, the relative contributions to the total test mass from the nucleon masses, electron mass and Coulomb energy are $\approx 1$, $\approx 2 \times 10^{-4}$ and $\approx 4 \times 10^{-3}$, respectively.

\section{Coherently oscillating classical dark-matter fields}
\label{Sec:ODM}

\subsection{Dark-matter theory}
\label{Sec:ODM_Theory}
Feebly interacting, low-mass (sub-eV) spinless particles are well-motivated candidates for DM. 
Perhaps the most renowned particle of this category is the canonical axion, which is a pseudoscalar (odd-parity) particle. 
Apart from the axion, low-mass scalar particles (such as the dilaton) may also exist in nature. 
Low-mass spinless particles can be produced non-thermally in the early Universe via the ``vacuum misalignment'' mechanism \cite{Vac_misal_A,Vac_misal_B,Vac_misal_C}, and they can subsequently form a coherently oscillating classical field \cite{Footnote2}:~$\phi \approx \phi_0 \cos(\omega_\phi t)$, where the angular frequency of oscillation is given by $\omega_\phi \approx m_\phi c^2/\hbar$, with $m_\phi$ being the mass of the spinless particle. 
Although these DM particles are typically produced with very small kinetic energies, they become virialised during the formation of galactic structures ($v_\textrm{vir} \sim 300~\textrm{km/s}$ locally), giving these particles the finite coherence time:~$\tau_\textrm{coh} \sim 2\pi / (m_\phi v_\textrm{vir}^2) \sim 10^6  t_\textrm{osc}$; i.e., $\Delta \omega_\phi / \omega_\phi \sim 10^{-6}$ (see \cite{Derevianko_DM-lineshape_2018} for details of the expected lineshape). 
In other words, the oscillations of this galactic DM field are practically monochromatic, with a quality factor of $Q_\phi \sim 10^6$. 
The oscillating DM field carries the non-zero time-averaged energy density:
\begin{equation}
\label{ODM_energy_density}
\left< \rho_\phi \right> \approx \rho_\phi \approx \frac{m_\phi^2 \phi_0^2}{2} \, , 
\end{equation}
and satisfies the non-relativistic equation of state $\left< p_\phi \right> \ll \left< \rho_\phi \right>$, making it an ideal candidate for cold DM. 
If spinless particles comprise the entirety of the observed DM, then their reduced de Broglie wavelength cannot exceed the DM halo size of the smallest dwarf galaxies ($R_\textrm{dwarf} \sim 1~\textrm{kpc}$). 
This places the following lower bound on their mass:~$m_\phi \gtrsim 10^{-22}~\textrm{eV}$, which can be relaxed if these particles make up only a sub-dominant fraction of the observed DM. 
In this section, we focus on the linear interactions of the field $\phi$ in (\ref{lin_portal}). 
We mention that one may also separately consider the case of quadratic interactions of the field $\phi$ in (\ref{quad_portal}), see \cite{Stadnik_2015_DM-LI,Stadnik_2015_DM-VFCs,Hees_2018_DM} for the various intricacies of such types of interactions.

\subsection{Michelson interferometers
~~~~~~~~~~~~~~~~~~~~~~~
(GEO\,600, Fermilab holometer)}
\label{Sec:ODM_Non_FP_exps}
Consider a power- and possibly signal-recycled (dual-recycled if both) laser interferometer without Fabry-Perot resonators in the two arms, as illustrated by the simplified layout in Fig.~\ref{Fig:GEO_config}. 
Archetypes of this Michelson configuration include the GEO\,600 interferometer ($L = 600~\textrm{m}$ without account of the single folding of the arms, $l = 8~\textrm{cm}$) and the Fermilab holometer ($L = 40~\textrm{m}$, $l = 1.3~\textrm{cm}$). 
The input laser beam is fed into the power recycling cavity consisting of the mirror `PRM' and the Michelson interferometer consisting of the beam-splitter `BS' and mirrors `ETMX' and `ETMY'. 
When operating at destructive interference at the dark port, the power recycling cavity enhances the circulating power, thus enhancing shot-noise-limited sensitivity. 
The signal recycling mirror `SRM' (if present) increases the low-frequency sensitivity of the Michelson interferometer. 
The inset in Fig.~\ref{Fig:GEO_config} shows how the laser beam traverses the beam-splitter. 

The interferometer output can be expressed in terms of the difference of the optical path lengths in the two arms of the interferometer, $\Delta L = L_x - L_y$, with $L_x \approx L_y$. 
DM-induced time-varying changes in the size of a freely-suspended beam-splitter of thickness $l$ will shift the main reflecting surface (power reflectivity of $R = 50\%$) by the amount $\delta l / 2$ in the frame of the interferometer. 
Assuming a nominal angle of the beam-splitter with respect to the interferometer arms of $45^\circ$, we have $\delta L_x \approx \delta [\sqrt{2} n l - l/ (2\sqrt{2}) - w/2]$, where for simplicity we have omitted a geometric correction factor from Snell's law of refraction, and $\delta L_y = -\delta l / (2\sqrt{2}) - \delta w/2$, where $w$ is the thickness of the freely-suspended arm mirrors `ETMX' and `ETMY' ($w = 10~\textrm{cm}$ for the GEO\,600 detector, while $w = 1.3~\textrm{cm}$ for the Fermilab holometer). 
Hence we have: 
\begin{equation}
\label{GEO600_delta-L1}
\delta \left( L_x - L_y \right) \approx \sqrt{2} (n \cdot \delta l +  l \cdot \delta n) \, . 
\end{equation}
Michelson interferometers typically do not use a usual reference cavity to stabilise the laser. 
Instead, the laser is stabilised to the common-mode interferometer arm length, which is isolated against length fluctuations (via the suspension points) via multiple-pendulum suspension systems for the mirrors. 
In this case, Eq.~(\ref{ri_VFCs_reference_cavity-suppressed}) applies for $\delta n$, while $\delta l$ is governed by Eq.~(\ref{steady-state_underdamped_solution}). 
Hence for a DM oscillation frequency $f_\textrm{DM}$ well below the frequency of the fundamental vibrational mode of the beam-splitter $f_{0,\textrm{BS}}$, we have: 
\begin{equation}
\label{GEO600_delta-L2A}
\frac{\delta \left( L_x - L_y \right)}{L} \approx \frac{\sqrt{2} n l}{L} \left( - \frac{\delta \alpha}{\alpha} - \frac{\delta m_e}{m_e}  \right) \, ,
\end{equation}
which is the case for the entire optimal frequency range of the GEO\,600 detector. 
On the other hand, for $f_\textrm{DM} \gg f_{0,\textrm{BS}}$, we have: 
\begin{equation}
\label{GEO600_delta-L2B}
\frac{\delta \left( L_x - L_y \right)}{L} \approx \frac{\sqrt{2} n l}{L} \left( -10^{-2} \frac{\delta \alpha}{\alpha} -5 \times 10^{-3} \frac{\delta m_e}{m_e}  \right) \, . 
\end{equation}
In the case of the Fermilab holometer, $f_{0,\textrm{BS}} \approx 200~\textrm{kHz}$. 
We note that the interferometer arm length $L=(L_x+L_y)/2$ can be eliminated from Eqs.~(\ref{GEO600_delta-L2A}) and (\ref{GEO600_delta-L2B}), but is included here to be consistent with the conventional strain calibration of the interferometer output. 

The sensitivity of measurements with a single interferometer (such as the GEO\,600 detector) to an oscillating DM field with finite coherence time $\tau_\textrm{coh}$ improves with the integration time $t_\textrm{int}$ as $\propto (t_\textrm{int})^{-1/2}$ in the temporally coherent regime $t_\textrm{int} \lesssim \tau_\textrm{coh}$, and then continues to improve as $\propto (t_\textrm{int} \tau_\textrm{coh})^{-1/4}$ in the temporally incoherent regime $t_\textrm{int} \gtrsim \tau_\textrm{coh}$. 
On the other hand, the sensitivity of cross-correlation measurements with a pair of independent and isolated interferometers (such as the two co-located interferometers of the Fermilab holometer) improves as $\propto (t_\textrm{int})^{-1/2}$ for all integration times, provided that the DM field is spatially coherent over the entire apparatus (meaning that two identical interferometers would see the same DM signal in this case). 
In the current Fermilab holometer setup, $L = 40~\textrm{m}$ and the spatial separation between the two independent interferometers is $0.9~\textrm{m}$. 
Since both of these length scales are much smaller than the reduced de Broglie wavelength of an oscillating DM field for the entire optimal frequency range of the Fermilab holometer, the DM field is therefore spatially coherent over the entire holometer.

In Michelson interferometers of the type discussed thus far, the DM effects on the beam-splitter give the main contribution to the output signal. 
Let us briefly also discuss some possible subleading contributions to the output signal. 
There are common-mode effects associated with temporal variations of the laser frequency and interferometer arm lengths. 
Since the two arms of an interferometer are practically equal in length, these common-mode effects are strongly suppressed in the output signal (and moreover the latter common-mode effect is further suppressed in the detector's optimal frequency range due to the multiple-pendulum suspension systems for the mirrors). 
An oscillating DM field also induces common-mode time-varying size changes of the arm end mirrors `ETMX' and `ETMY', which cancel for a homogeneous DM field $\phi = \phi_0 \cos(m_\phi t)$ if these mirrors have the same thickness. 
In reality, there is a small non-zero contribution to the output signal due to a phase difference in the oscillating DM field between the two arm end mirrors, since in the laboratory frame of reference an oscillating DM field contains a position-dependent term:~$\phi = \phi_0 \cos(m_\phi t - \v{p}_\phi \cdot \v{r})$, where $\v{p}_\phi \approx m_\phi \left<\v{v}_\phi\right>$ is the average momentum of a DM particle as seen in the laboratory frame. 
In this case, there is an effect suppressed by $\sim L/\lambdabar_\textrm{dB} \lesssim |\v{v}_\phi| \sim 10^{-3}$ in the optimal frequency range of ground-based detectors. 
Incidentally, since $l , w \ll L$, this also justifies treating the beam-splitter and test masses as pointlike objects. 

\begin{figure*}[h!]
\includegraphics[width=0.48\linewidth]{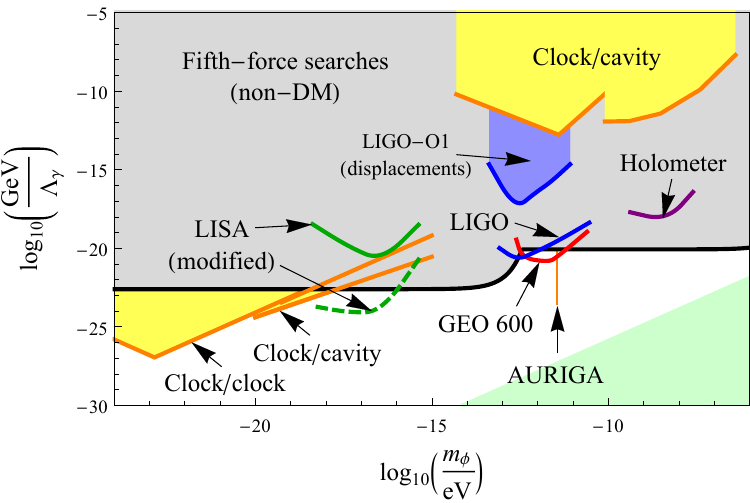}
\hspace{5mm}
\includegraphics[width=0.48\linewidth]{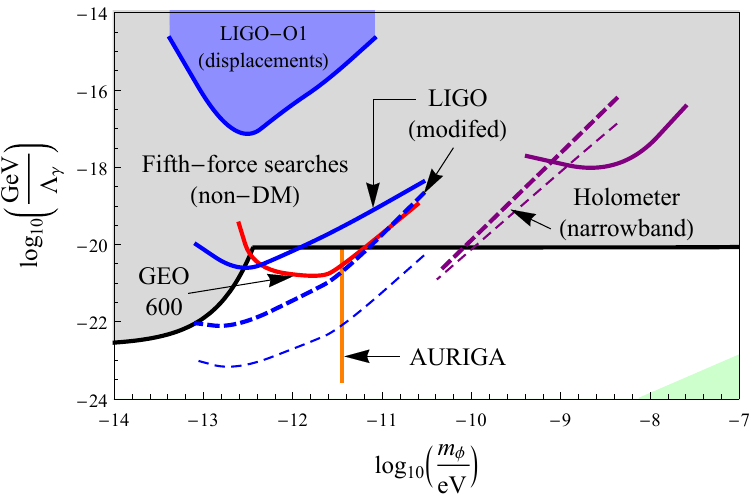}
\par\bigskip
\includegraphics[width=0.48\linewidth]{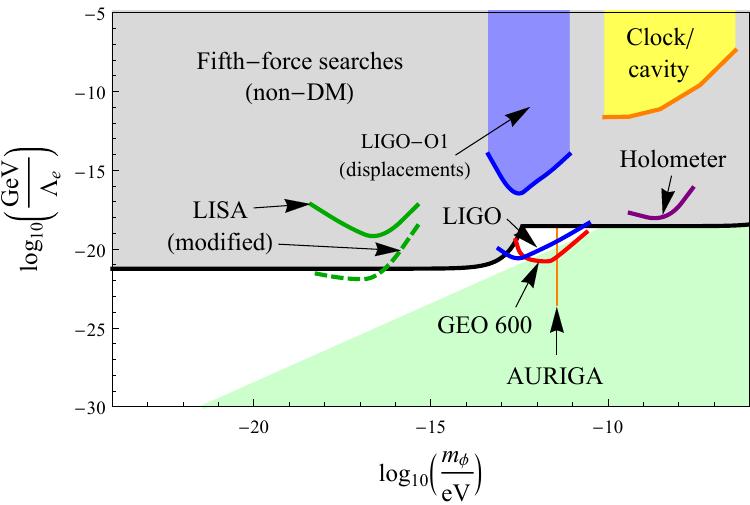}
\hspace{5mm}
\includegraphics[width=0.48\linewidth]{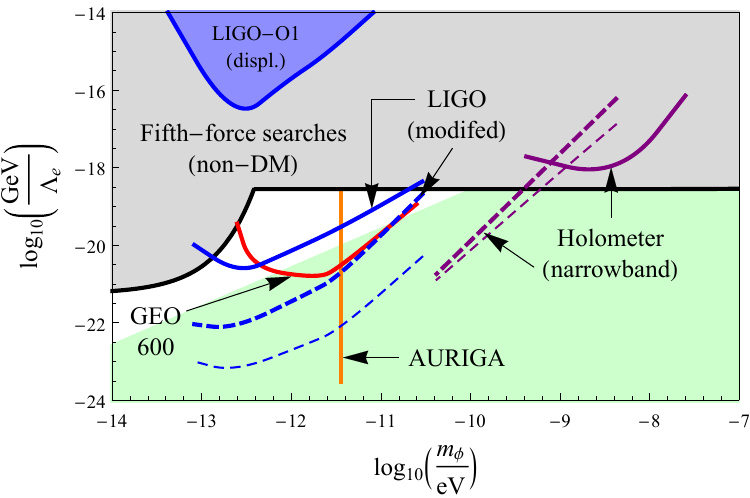}
\par\bigskip
\centering
\includegraphics[width=0.48\linewidth]{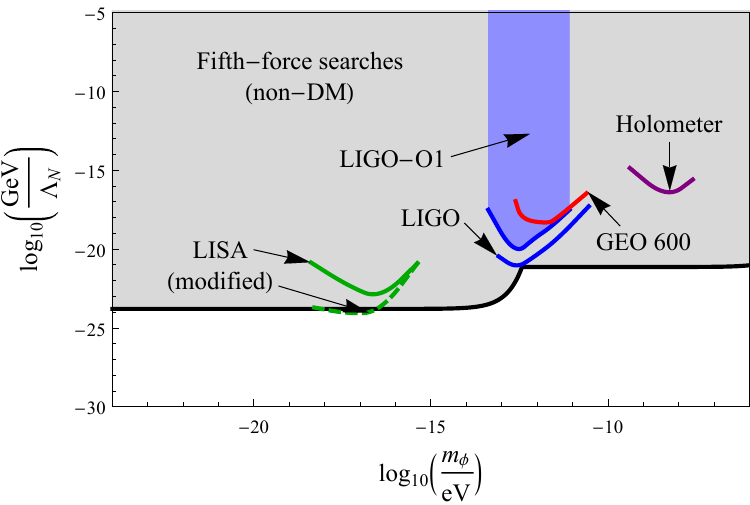}
\hspace{5mm}
\includegraphics[width=0.48\linewidth]{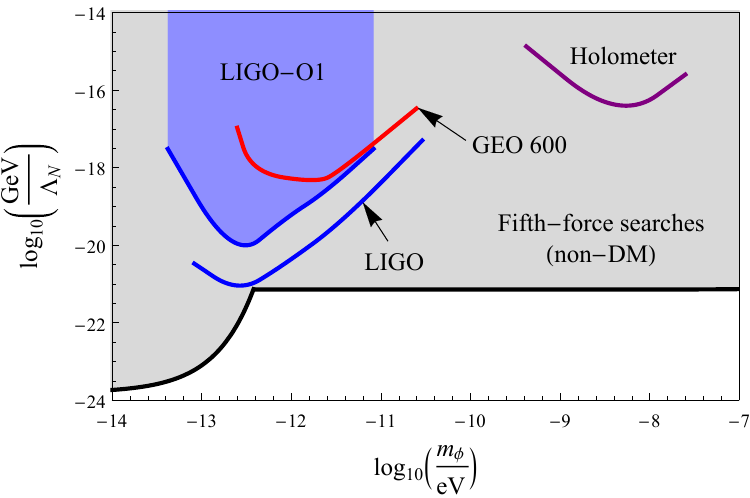}
\caption{(Color online). 
From top to bottom:~Physical parameter spaces for the linear interactions of an oscillating DM field $\phi$ with the electromagnetic field (photon), electron and nucleons, as functions of the DM particle mass $m_\phi$. 
The solid lines denote the estimated sensitivities of current ground-based laser interferometers (red = GEO\,600, blue = LIGO, purple = Fermilab holometer). 
The dashed blue line denotes the projected sensitivity of a single modified LIGO interferometer, in which the thicknesses of the Fabry-Perot mirrors in one of the interferometer arms are changed by $10\%$, and operating at the design sensitivity of Advanced LIGO, while the thin dashed blue line denotes the analogous sensitivity for a pair of modified LIGO interferometers. 
The dashed purple line denotes the estimated sensitivity of a single small-scale Michelson interferometer operating in the resonant narrowband regime near room temperature, with $Q \sim 10^6$ and covering a DM particle mass range of $\Delta m_\phi/m_\phi \approx 1$, while the thin dashed purple line denotes the analogous sensitivity for a pair of co-located interferometers. 
The solid green line denotes the projected sensitivity of the space-based LISA interferometer in its standard configuration, while the dashed green line denotes the projected sensitivity of LISA with some of its Au-Pt alloy test masses replaced by Be test masses. 
All of these sensitivities assume a total integration time of $t_\textrm{int} \sim 10^8~\textrm{s}$ and saturation of the average local cold DM density of $\rho_\textrm{DM} \approx 0.4~\textrm{GeV}/\textrm{cm}^3$. 
The region in grey denotes existing non-DM-based constraints from fifth-force experiments \cite{MICROSCOPE_2017,MICROSCOPE_2018,Torsion_balance_Uranium_1999,Stadnik_2016_5F}. 
The regions in yellow denote existing DM-based constraints from experiments involving clock-clock comparisons \cite{Leefer_2015_Dy-DM,Hees_2016_clock-DM}, clock-cavity comparisons \cite{Ye_2018_cavity-DM,Aharony_2019_cavity-DM,Savalle_2019_cavity-DM,Tretiak_2019_cavity-DM} and the AURIGA resonant-bar detector \cite{AURIGA_2017_DM}. 
The region in blue denotes constraints derived in the present work from the consideration of time-varying centre-of-mass displacements of LIGO's test masses in the LIGO-O1 data using the results of data analysis taken from \cite{GW-Dark_photon_2019}. 
The region in pale green represents the region of parameter space that is technically natural for a new-physics cut-off scale of $\Lambda \sim 10~\textrm{TeV}$. 
}
\label{Fig:ODM_sensitivities}
\end{figure*}

Using Eqs.~(\ref{lin_VFCs}), (\ref{ODM_energy_density}), (\ref{GEO600_delta-L2A}) and (\ref{GEO600_delta-L2B}), we estimate the current sensitivities of GEO\,600 \cite{GEO600_2013,GEO600_2015} and the Fermilab holometer using both of its co-located interferometers \cite{Fermilab_holometer_2016,Fermilab_holometer_2017} to the linear interactions of the DM field $\phi$ with the photon and electron in (\ref{lin_portal}). 
We present these estimates as solid lines in Fig.~\ref{Fig:ODM_sensitivities} (red = GEO\,600, purple = Fermilab holometer using both of its co-located interferometers), assuming that the $\phi$ particles saturate the average local cold DM density of $\rho_\textrm{DM} \approx 0.4~\textrm{GeV}/\textrm{cm}^3$. 
We note that, because the amplitude and effects of the oscillating DM field scale as $\propto 1/m_\phi$, the sensitivities of the detectors to the underlying DM interactions in (\ref{lin_portal}) peak at frequencies somewhat lower than the optimal frequencies for characteristic strains.

\subsection{Fabry-Perot-Michelson interferometers
~~~~~~~~~~~~~~~~~~~~~~~
(LIGO, VIRGO, KAGRA)}
\label{Sec:ODM_FP_exps}
Consider now a dual-recycled Fabry-Perot-Michelson interferometer of the type shown in the simplified layout in Fig.~\ref{Fig:LIGO_config}. 
Archetypes of this configuration include the LIGO ($L = 4~\textrm{km}$, $l = 6~\textrm{cm}$), VIRGO ($L = 3~\textrm{km}$, $l = 5.5~\textrm{cm}$) and KAGRA ($L = 3~\textrm{km}$, $l = 8~\textrm{cm}$) interferometers. 
In contrast to the layout of a dual-recycled Michelson interferometer (see Fig.~\ref{Fig:GEO_config}), the additional mirrors `ITMX' and `ITMY' form Fabry-Perot resonators in each arm. 
While these Fabry-Perot cavities increase the strain sensitivity of the interferometer to gravitational waves, they generally reduce the effective sensitivity of the interferometer to DM effects on the beam-splitter, 
since in this case for each to-and-back passage across the beam-splitter, the laser beam encounters $N \gg 1$ to-and-back passages within the Fabry-Perot cavities. 
In this case, Eq.~(\ref{GEO600_delta-L1}) is modified accordingly: 
\begin{equation}
\label{LIGO_delta-L1}
\delta \left( L_x - L_y \right)_\textrm{eff} \approx \frac{\sqrt{2}}{N_\textrm{eff}} (n \cdot \delta l +  l \cdot \delta n) \, . 
\end{equation}
The bandwidth of the Fabry-Perot-Michelson interferometer is determined by the total propagation time of
photons through the Fabry-Perot arm cavities, but is increased (in the case of LIGO, VIRGO and KAGRA) by
the signal recycling mirror (which in this configuration effectively shortens the number of round-trips
for those photons that carry signal information in the measurement band).
For DM oscillation frequencies below this modified bandwidth of the Fabry-Perot-Michelson interferometer, we have $1/N_\textrm{eff} \approx 1/N$. 
However, for DM oscillation frequencies above the modified bandwidth of the Fabry-Perot-Michelson interferometer, $1/N_\textrm{eff}$ increases approximately linearly with $f_\textrm{DM}$ due to the finite propagation speed of photons. 
If the thicknesses of the freely-suspended Fabry-Perot arm mirrors in both arms are equal (and comparable to the beam-splitter thickness), then the quasi-common-mode effects of an oscillating DM field on the arm mirrors, as discussed in Sec.~\ref{Sec:ODM_Non_FP_exps}, are subleading in the optimal frequency range of the detector, provided that $N \lesssim 10^3$. 
In the current LIGO setup, $l = 6~\textrm{cm}$, $w = 20~\textrm{cm}$ and $N \sim 10^2$, so the beam-splitter effect in Eq.~(\ref{LIGO_delta-L1}) indeed gives the main contribution to the output signal in this case; 
however, the quasi-common-mode effects on the arm mirrors in this case are not as strongly suppressed (compared to the beam-splitter effect) as they are in the case of a Michelson interferometer (see Sec.~\ref{Sec:ODM_Non_FP_exps}).

While some Fabry-Perot-Michelson interferometers use a small-scale reference cavity to initially lock the laser, the laser is ultimately locked and stabilised to the common-mode interferometer arm length, which is isolated against length fluctuations (via the suspension points) via multiple-pendulum suspension systems for the mirrors. 
In this case, Eq.~(\ref{ri_VFCs_reference_cavity-suppressed}) applies for $\delta n$. 
The entire optimal frequency range of the LIGO detector lies well below the frequency of the fundamental vibrational mode of the beam-splitter, and so $\delta l$ is governed by the adiabatic formula (\ref{size_changes1}). 
Thus, we have: 
\begin{equation}
\label{LIGO_delta-L2B}
\frac{\delta \left( L_x - L_y \right)_\textrm{eff}}{L} \approx \frac{\sqrt{2} n l}{N_\textrm{eff} L} \left( - \frac{\delta \alpha}{\alpha} - \frac{\delta m_e}{m_e}  \right) \, . 
\end{equation}

If the thicknesses of mirrors `ETMX' and `ITMX' differ by an amount $\Delta w$ with respect to the thicknesses of mirrors `ETMY' and `ITMY', then there will be an additional contribution to the output signal given by: 
\begin{equation}
\label{LIGO-modified_delta-L2}
\frac{\delta \left( L_x - L_y \right)}{L} \approx - \frac{\Delta w}{L} \left( - \frac{\delta \alpha}{\alpha} - \frac{\delta m_e}{m_e} \right) \, . 
\end{equation}
In the current LIGO interferometers, $|\Delta w| \approx 80~\mathrm{\upmu}\textrm{m}$, and so in this case the contribution in (\ref{LIGO-modified_delta-L2}) will be smaller than the DM effect on the beam-splitter in Eq.~(\ref{LIGO_delta-L2B}) for the entire optimal frequency range of the LIGO detector. 
However, it is possible to significantly increase the sensitivity of a Fabry-Perot-Michelson interferometer to time-varying $\alpha$ and $m_e$ by making the thicknesses of the freely-suspended Fabry-Perot arm mirrors sufficiently different in the two arms. 

Using Eqs.~(\ref{lin_VFCs}), (\ref{ODM_energy_density}), (\ref{LIGO_delta-L2B}) and (\ref{LIGO-modified_delta-L2}), we estimate the sensitivities of LIGO \cite{aLIGO_2016} in both its current configuration and in a modified configuration where the arm mirrors in the two arms have appreciably different thicknesses (we take $\Delta w / w = 10\%$ for concreteness) to the linear interactions of the DM field $\phi$ with the photon and electron in (\ref{lin_portal}). 
We present these estimates as blue lines in Fig.~\ref{Fig:ODM_sensitivities} (solid line = current configuration and single detector, dashed line = modified configuration and single detector, thin dashed line = modified configuration and both detectors \cite{Footnote_LIGO_CC}), assuming that the $\phi$ particles saturate the average local cold DM density. 
Note the difference in the shapes of the curves for single detectors at higher frequencies, due to the dominant effect from the arm mirrors in the modified case, compared to the case where the beam-splitter effect dominates. 
The sensitivity of cross-correlation measurements using both LIGO detectors is enhanced compared to measurements using a single LIGO detector by the factor $\sim \left( t_\textrm{int} / \tau_\textrm{coh} \right)^{1/4} \gg 1$. 
Laser interferometers of different types (e.g., one Michelson interferometer and one Fabry-Perot-Michelson interferometer) can also benefit significantly from cross-correlation measurements, provided that their individual sensitivities to the underlying DM interaction parameters are similar.

\subsection{Resonant narrowband experiments}
\label{Sec:ODM_resonance-narrowband_exps}
In Secs.~\ref{Sec:ODM_Non_FP_exps} amd \ref{Sec:ODM_FP_exps}, we considered broadband detection strategies. 
In this section, we consider the possibility of resonant narrowband searches with laser interferometers. 
The crucial observation is that the oscillations of the galactic DM field are expected to be practically monochromatic, with a quality factor of $Q_\phi \sim 10^6$. 
We begin with small-scale Michelson interferometers, having in mind the Fermilab holometer as a possible platform. 
If the DM oscillation frequency matches the fundamental frequency of the longitudinal vibrational mode of the beam-splitter ($\approx 200~\textrm{kHz}$ for the Fermilab holometer), then, according to Eq.~(\ref{steady-state_underdamped_solution}), the DM-induced time-varying size changes of the beam-splitter will be enhanced by the factor $Q = \textrm{min} \{ Q_\phi , Q_\textrm{mech} \}$: 
\begin{equation}
\label{Fermilab_delta-L2_resonance}
\frac{\delta \left( L_x - L_y \right)}{L} \approx \frac{\sqrt{2} Q n l}{L} \left( - \frac{\delta \alpha}{\alpha} - \frac{\delta m_e}{m_e} \right) \, . 
\end{equation}
Materials with quality factors comparable to the DM quality factor of $Q_\phi \sim 10^6$ are available. 
However, in order to achieve the desired mechanical quality factor of at least $Q_\textrm{mech} \sim 10^6$ (and hence an overall quality factor of $Q \sim 10^6$), one would need to ensure that the clamps which support the beam-splitter are designed in such a way as not to degrade the overall quality factor. 

For simplicity, we neglect higher-order-harmonic vibrational modes in the ensuing discussion and, furthermore, assume that the measurements are limited by Brownian thermal noise. 
By the equipartition theorem, the potential energy associated with the longitudinal vibrational mode of the beam-splitter is given by $M_\textrm{BS} \omega_{0,\textrm{BS}}^2 \left< x^2 \right>/2 = k_\textrm{B} T/2$, where $\left< x^2 \right>$ is the mean-square displacement of the reflecting surface, $T$ is the temperature and $k_\textrm{B}$ is the Boltzmann constant. 
In the vicinity of the longitudinal vibrational mode resonance, the thermal-noise amplitude spectral density is hence given by: 
\begin{equation}
\label{Brownian_thermal_noise}
\textrm{ASD (thermal noise)} = \sqrt{ \frac{Q k_\textrm{B} T}{2 \pi M_\textrm{BS} f_{0,\textrm{BS}}^3} } \, . 
\end{equation}
We thus see that beam-splitters of larger transverse sizes are advantageous with regards to thermal noise (in the current Fermilab holometer setup, $M_\textrm{BS} = 130~\textrm{g}$). 
By cooling the system from room temperature to liquid-helium temperature, the thermal noise can be reduced by a factor of $\sim 10$. 
Brownian thermal noise, being broadband, scales as $\propto Q^{1/2}$, and so the signal-to-noise ratio in such resonant narrowband experiments scales as $\propto Q^{1/2}$. 

In order to scan over a range of different DM particle masses, one must incrementally vary the fundamental resonance frequency. 
One can alter $f_{0,\textrm{BS}}$ by changing the thickness of the beam-splitter via ablation and polishing. 
For very small incrementations of $f_{0,\textrm{BS}}$, it would generally be more efficient to incrementally change the temperature of the system. 
Near room temperature, silica has a thermal expansion coefficient of $\sim 10^{-6}~\textrm{K}^{-1}$ and a sound speed of $v_s \approx 6~\textrm{km/s}$, while sapphire has a thermal expansion coefficient of $\sim 10^{-5}~\textrm{K}^{-1}$ and a sound speed of $v_s \approx 10~\textrm{km/s}$. 
The relative sound speed change in silica is $\sim 10^{-3}$ over a temperature interval of $\Delta T \sim 10~\textrm{K}$ near room temperature, which would cover a frequency range of $\Delta f_{0,\textrm{BS}}/f_{0,\textrm{BS}} \sim 10^{-3}$ over the same temperature interval, see Eq.~(\ref{fund_freq_long_mode}). 
Beam-splitter thicknesses in the range $\sim 1 - 10~\textrm{cm}$ are routinely used in existing Michelson interferometers. 
In dedicated experiments, beam-splitter thicknesses in the range $\textrm{few~mm} - 30~\textrm{cm}$ may be achievable. 
Thus, fundamental frequencies of the beam-splitter in the range $\sim 10~\textrm{kHz} - 1~\textrm{MHz}$ can reasonably be covered. 
This frequency range overlaps with the frequency range of proposed resonant narrowband experiments with resonant-mass detectors \cite{Arvanitaki2016_resonance-bar}. 

Just like the broadband searches discussed in Secs.~\ref{Sec:ODM_Non_FP_exps} amd \ref{Sec:ODM_FP_exps}, the sensitivity of such narrowband searches with a single Michelson interferometer improves with the integration time as $\propto (t_\textrm{int})^{-1/2}$ for $t_\textrm{int} \lesssim \tau_\textrm{coh}$, then as $\propto (t_\textrm{int} \tau_\textrm{coh})^{-1/4}$ for $t_\textrm{int} \gtrsim \tau_\textrm{coh}$, while the sensitivity of these narrowband searches with a pair of co-located Michelson interferometers improves as $\propto (t_\textrm{int})^{-1/2}$ for all integration times. 
Using Eqs.~(\ref{lin_VFCs}), (\ref{ODM_energy_density}), (\ref{Fermilab_delta-L2_resonance}) and (\ref{Brownian_thermal_noise}), we estimate the sensitivities of a single small-scale Michelson interferometer and a pair of co-located small-scale Michelson interferometers using the above narrowband approach and operating near room temperature to the linear interactions of the DM field $\phi$ with the photon and electron in (\ref{lin_portal}), assuming that the measurements are limited by Brownian thermal noise and that all of the dimensions of the beam-splitter are altered in a proportional manner. 
We present these estimates as the dashed and thin dashed purple lines, respectively, in Fig.~\ref{Fig:ODM_sensitivities}, for $Q \sim 10^6$, covering a DM particle mass range of $\Delta m_\phi/m_\phi \approx 1$ and assuming that the $\phi$ particles saturate the average local cold DM density. 
We note that, because the amplitude of the DM field scales as $\propto 1/m_\phi$ and that the resonant condition $f_\textrm{DM} = f_{0,\textrm{BS}}$ implies the scaling relation $l \propto 1/m_\phi$ for a fixed sound speed [see Eq.~(\ref{fund_freq_long_mode})], the size of the resonantly-enhanced DM effects in our proposed narrowband experiments scales roughly as $\propto 1/m_\phi^2$. 
This scaling strongly favours such narrowband searches at lower DM particle masses. 

As mentioned in Sec.~\ref{Sec:ODM_FP_exps}, some Fabry-Perot-Michelson interferometers use reference cavities to initially lock the laser. 
In such cases, there can be a similar resonant enhancement when the DM oscillation frequency matches the fundamental frequency of the longitudinal vibrational mode of the reference cavity. 
For a typical reference cavity length of $L_\textrm{cav} \sim 0.5~\textrm{m}$ and most commonly used materials, the fundamental frequency of the longitudinal vibrational mode within the cavity is $f_{0,\textrm{cav}} \sim 6~\textrm{kHz}$, see Eq.~(\ref{fund_freq_long_mode}). 
Since it is generally preferable to lock and stabilise the laser to the common-mode interferometer arm length, rather than to a small-scale reference cavity, it seems difficult to take advantage of this resonant enhancement factor in practice for most Fabry-Perot-Michelson interferometers.
It may be possible though for the VIRGO interferometer, where the reference cavity is suspended within the main vacuum envelope and thus is in a lower-noise environment than the reference cavity in LIGO. 
For this method to work, the read-out of the reference cavity length would have to be sufficiently sensitive. 

One may alternatively modify broadband clock-cavity comparison experiments of the type considered in \cite{Stadnik_2016_cavities,Wcislo_2016_cavity-DM,Ye_2018_cavity-DM,Aharony_2019_cavity-DM,Tretiak_2019_cavity-DM} to scan over a range of DM particle masses where the reference cavity length changes are resonantly enhanced. 
In this case, we would have \cite{Footnote4}: 
\begin{equation}
\label{clock-cavity_resonance}
\frac{\delta \left( \omega_\textrm{cavity} /  \omega_\textrm{atom} \right)}{\omega_\textrm{cavity} /  \omega_\textrm{atom}} \approx Q \left( \frac{\delta \alpha}{\alpha} + \frac{\delta m_e}{m_e} \right) \, , 
\end{equation}
where we have assumed that the relative sensitivity coefficients of the atomic transition frequency $\omega_\textrm{atom}$ to apparent changes in $\alpha$ and $m_e$ are much smaller than $Q$. 
Most reference cavities have quality factors well in excess of the DM quality factor of $Q_\phi \sim 10^6$, giving an overall quality factor of $Q \sim 10^6$ in this case. 
Reference cavities with lengths in the range $\sim 0.1 - 1~\textrm{m}$ are readily available, meaning that fundamental frequencies of the cavity in the range $\sim 3 - 30~\textrm{kHz}$ can reasonably be covered.

\subsection{Space-based experiments (LISA)}
\label{Sec:ODM_space_exps}
Let us briefly discuss space-based laser interferometry experiments. 
The archetypal example in this case is LISA, which is a three-arm interferometer in a triangular geometry ($L = 2.5 \times 10^{9}~\textrm{m}$). 
On board each of the three spacecrafts is a pair of quasi-freely-floating cubic test masses of side length $s = 4.6~\textrm{cm}$. 
At the low DM oscillation frequencies that lie in the optimal frequency range of the LISA detector ($\sim 10^{-4} - 10^{-1}~\textrm{Hz}$), the test cube size changes are described by the adiabatic formula (\ref{size_changes1}): 
\begin{align}
\label{size_changes_LISA-cubes}
\frac{\delta s}{s} &\approx - \frac{\delta \alpha}{\alpha} - \frac{\delta m_e}{m_e} \notag \\ 
&\approx - \frac{\sqrt{2 \rho_\phi}}{m_\phi} \cos(m_\phi t) \left( \frac{1}{\Lambda_\gamma} + \frac{1}{\Lambda_e} \right)  \, , 
\end{align}
where we have used Eqs.~(\ref{lin_VFCs}) and (\ref{ODM_energy_density}) in the second line. 
Additionally, the freely-floating Au-Pt test cubes will experience a time-varying acceleration in accordance with Eq.~(\ref{test_mass_acceleration_general}). 
For the oscillating DM field $\phi = \phi_0 \cos(m_\phi t - \v{p}_\phi \cdot \v{r})$, we find: 
\begin{align}
\label{test_mass_acceleration_LISA}
\delta \v{a}_\textrm{test} \approx &- \sqrt{2 \rho_\phi} \v{v}_\phi \sin(m_\phi t)   \notag \\
&\times \left( \frac{4 \times 10^{-3}}{\Lambda_\gamma} + \frac{2 \times 10^{-4}}{\Lambda_e} + \frac{1}{\Lambda_N} \right)  \, , 
\end{align}
where we have again used Eqs.~(\ref{lin_VFCs}) and (\ref{ODM_energy_density}). 
The freely-floating test masses will, therefore, undergo the following time-varying centre-of-mass displacements: 
\begin{align}
\label{test_mass_displacement_LISA}
\delta \v{x}_\textrm{test} \approx &+ \frac{\sqrt{2 \rho_\phi} \v{v}_\phi}{m_\phi^2} \sin(m_\phi t)   \notag \\
&\times \left( \frac{4 \times 10^{-3}}{\Lambda_\gamma} + \frac{2 \times 10^{-4}}{\Lambda_e} + \frac{1}{\Lambda_N} \right)  \, . 
\end{align}
We note that the time-varying size changes and centre-of-mass displacements of the test cubes in Eqs.~(\ref{size_changes_LISA-cubes}) and (\ref{test_mass_displacement_LISA}), respectively, are out of phase with respect to each other by the factor $\pi/2$, and that the former effect scales as $\propto 1/m_\phi$, while the latter scales as $\propto 1/m_\phi^2$. 
Additionally, in contrast to the DM-induced time-varying size changes of the test cubes, the DM-induced time-varying centre-of-mass displacements of the test cubes are anisotropic, meaning that the resulting observable signatures will strongly depend on the orientation of the detector and its components with respect to $\v{v}_\phi$. 

The LISA interferometer operates on the principle of time-delay interferometry, which basically involves measuring a particular linear combination of the three arm lengths to cancel the laser phase noises that would otherwise be imprinted in length measurements of the unequal (and naturally time-varying) arm lengths of the interferometer. 
Hence the common-mode effects of a homogeneous DM field $\phi = \phi_0 \cos(m_\phi t)$ on test masses in different spacecrafts will cancel to leading order. 
In particular, centre-of-mass displacements of identical test masses by a homogeneous DM field correspond to the translation of the entire system. 
The leading non-vanishing contribution to the output signal arises due to phase differences in the oscillating DM field, $\phi = \phi_0 \cos(m_\phi t - \v{p}_\phi \cdot \v{r})$, between pairs of spacecraft, and so the DM-induced effects on the pairs of test masses in Eqs.~(\ref{size_changes_LISA-cubes}) and (\ref{test_mass_displacement_LISA}) will be effectively suppressed by the factor $\sim L/\lambdabar_\textrm{dB} \ll 1$. 
It is possible to significantly increase the sensitivity of LISA to time-varying $\alpha$ and $m_e$ by replacing some of the freely-floating Au-Pt alloy test masses by test masses made of much lighter elements (such as Be, Al and/or Ti) \cite{Footnote4b}. 
In this case, some of the common-mode suppression would be lifted due to the maximally different mass-energy contributions of elements from different regions of the periodic table, see Eq.~(\ref{Bethe-Weizsaecker_formula}). 
Using Eqs.~(\ref{size_changes_LISA-cubes}) and (\ref{test_mass_displacement_LISA}), we estimate the projected sensitivities of LISA \cite{LISA_2017} in both its standard configuration and with our suggested modification (we suppose that some of the Au-Pt alloy test masses are replaced by Be test masses for concreteness) to the linear interactions of the DM field $\phi$ with the photon, electron and nucleons in (\ref{lin_portal}). 
We present these estimates as green lines in Fig.~\ref{Fig:ODM_sensitivities} (solid line = standard configuration, dashed line = modified configuration), assuming that the $\phi$ particles saturate the average local cold DM density.

In space-based laser-interferometric detectors, the time-varying centre-of-mass displacements of the test masses are generally more important than the time-varying size changes of the test masses. 
Indeed, the ratio of the two effects, modulo different material-dependent sensitivity coefficients, is of the order $\sim v_\phi/(m_\phi s)$, which is $\gg 1$ in the optimal frequency range of the LISA detector. 
In contrast, in ground-based Michelson interferometers, the (common-mode-suppressed) time-varying centre-of-mass displacements of (identical) test masses and beam-splitter are generally less important than time-varying size changes of the beam-splitter. 
In this case, the ratio of the two effects, modulo different system/material-dependent sensitivity coefficients and geometric factors, is of the order $\sim v_\phi^2 L / l$, which is $\ll 1$ for a typical ground-based detector \cite{Footnote5}. 
Using Eq.~(\ref{test_mass_displacement_LISA}), together with the relevant sensitivity coefficients for silica test masses presented following Eq.~(\ref{Bethe-Weizsaecker_formula}), we estimate the current sensitivities of ground-based laser interferometers to the linear interaction of the DM field $\phi$ with nucleons in (\ref{lin_portal}). 
We present these estimates as solid lines in Fig.~\ref{Fig:ODM_sensitivities} (red = GEO\,600, blue = LIGO, purple = Fermilab holometer using both of its co-located interferometers), assuming that the $\phi$ particles saturate the average local cold DM density. 
Using the results of the recent data analysis in Ref.~\cite{GW-Dark_photon_2019} that searched for analogous time-varying centre-of-mass displacements of LIGO's test masses in the LIGO-O1 data due to dark-photon interactions (instead of scalar interactions), we place bounds on the linear interactions of the DM field $\phi$ with the photon, electron and nucleons in (\ref{lin_portal}). 
These LIGO-O1 bounds are denoted by the blue region in Fig.~\ref{Fig:ODM_sensitivities}. 
We note that, in contrast to the beam-splitter effect in Eq.~(\ref{LIGO_delta-L2B}), there is no $1/N_\textrm{eff}$ suppression factor in Fabry-Perot-Michelson interferometry searches for time-varying centre-of-mass displacements of the beam-splitter and test masses. 
We also note that DM-induced time-varying centre-of-mass displacements of freely-suspended interferometer components are phenomenologically more interesting for dark-photon interactions \cite{GW-Dark_photon_2019,GW-Dark_photon_2018}, due to the lack of an extra velocity suppression factor $v_\phi \ll 1$ compared with the scalar interactions considered in the present work [see Eqs.~(\ref{test_mass_acceleration_LISA}) and (\ref{test_mass_displacement_LISA}), as well as \cite{Arvanitaki2015_dilaton-GW,Suyama2015_dilaton-GW}].

\subsection{Local dark-matter overdensities}
\label{Sec:local_DM_overdensities}
For DM particle masses corresponding to the optimal frequency ranges of ground-based laser interferometers, it is possible for the DM density near the surface of Earth to be many orders of magnitude greater than the average local cold DM density of $\rho_\textrm{DM} \approx 0.4~\textrm{GeV}/\textrm{cm}^3$ inferred from galactic rotation curve measurements for our Galaxy. 
Such a situation may arise, e.g., due to the capture of an overdense region of DM by the gravitational well of Earth or the Sun. 
This is in stark contrast to laboratory experiments that search for DM with particle masses $m_\phi \sim 10^{-22}~\textrm{eV}$ (see, e.g., \cite{Leefer_2015_Dy-DM,Hees_2016_clock-DM,nEDM_2017_axion-DM}), where the Heisenberg uncertainty principle prevents the gravitational collapse of such ultra-low-mass DM fields on length scales shorter than their reduced de Broglie wavelength (which is astronomical in this case). 
We focus on two specific cases of static and uniformly-distributed overdensities of DM that are centered on the Sun and Earth, respectively. 

For a spherical DM overdensity centered on the Sun, the most stringent bounds on the largest allowable DM density near Earth's surface generally come from planetary ephemeris measurements. 
For a spherical DM overdensity of radius $R \approx 1~\textrm{AU}$, the largest allowable DM density near Earth's surface is $\sim 10^5$ times the average local cold DM density \cite{Pitjeva_2013_DM-overdensity}. 
In this case, the DM particles are assumed to be gravitationally bound to the Sun. 
By the virial theorem, $\left< v_\phi^2 \right> \approx G M_\odot / R \sim 10^{-8}$ at the position of Earth. 
Since the DM particles are also assumed to be localised within the sphere of radius $R$, we further require that $\lambdabar_\textrm{dB} \lesssim R$, which sets the requirement $m_\phi \gtrsim 10^{-14}~\textrm{eV}$. 
This is serendipitous, because this includes the entire optimal frequency ranges of current ground-based laser interferometers. 
Additionally, the quality factor associated with the DM oscillations is $Q_\phi \sim 10^8$ in this case, which is $\sim 100$ times larger than in the usual ``galactic picture''. 

For a spherical DM overdensity centered on Earth, the most stringent bounds on the largest allowable DM density near Earth's surface generally come from a combination of lunar laser ranging and geodetic surveyance measurements. 
For a spherical DM overdensity of radius $R \approx 60 R_\oplus$, the largest allowable DM density near Earth's surface is $\sim 10^{11}$ times the average local cold DM density \cite{Adler_2008_DM-overdensity}. 
In this case, the DM particles are assumed to be gravitationally bound to Earth. 
By the virial theorem, $\left< v_\phi^2 \right> \approx G M_\oplus / R_\oplus \sim 10^{-9}$ near Earth's surface. 
Since the DM particles are assumed to be localised within the sphere of radius $R$, we require that $\lambdabar_\textrm{dB} \lesssim R$, which sets the requirement $m_\phi \gtrsim 10^{-11}~\textrm{eV}$. 
This is fortunate, as this includes the entire optimal frequency range of the current Fermilab holometer. 
In this case, the quality factor associated with the DM oscillations is $Q_\phi \sim 10^9$, which is $\sim 10^3$ times larger than in the usual galactic picture. 

We note that the sensitivities of ground-based laser interferometers to the underlying DM interactions are enhanced with respect to non-DM-based experiments for the scenarios discussed in this section, not only because of the increased DM density (the size of the DM effects scale as $\propto \sqrt{\rho_\phi}$ for linear interactions, while non-DM effects are independent of $\rho_\phi$), but also because of the increased coherence time $\tau_\textrm{coh} \propto Q_\phi$ (which diminishes the role of incoherent averaging for measurements with a single interferometer when $t_\textrm{int} \gtrsim \tau_\textrm{coh}$ and, in the case of resonant narrowband experiments, may increase the signal-to-noise ratio by increasing the overall quality factor $Q$). 
Another possible way for the DM density near the surface of Earth to be greatly enhanced is via the formation and subsequent capture of DM objects that are bound by their own self-gravity and self-interactions \cite{Eby_DM-star_2019}.

\section{Topological defects}
\label{Sec:TDM}

\subsection{Theory of topological defects}
\label{Sec:TDM_Theory}
Topological defects are stable solitonic configurations of DM fields that may be produced as a result of a phase transition in the early Universe \cite{Vilenkin_1985_TDs}. 
These (possibly macroscopic) objects may come in a variety of dimensionalities:~0D (monopoles), 1D (strings) or 2D (domain walls). 
As a simple illustrative example, consider a real spinless field $\phi$ in one spatial dimension with the self-potential $V(\phi) = \sigma (\phi^2 - \eta^2)^2$, which has two energetically equivalent vacua at $\phi = +\eta$ and $\phi = -\eta$. 
In this case, a stable domain wall with the transverse profile $\phi(x) = \eta \tanh(m_\phi x)$, where $m_\phi = \sqrt{\sigma} \eta$, will form between the two vacua. 
The transverse size of this domain wall is set by the reduced Compton wavelength of the underlying field, $d \sim 1/m_\phi$, and any physical effects produced by this wall arise only at the boundary between the two vacua. 
The energy density stored inside a domain wall is given by $\rho_\textrm{inside} \sim m_\phi^2 \phi_m^2 \sim \phi_m^2/d^2$, where $\phi_m$ is the amplitude (relative to the vacuum states) of the field $\phi$ inside the wall. 
A network of finite-sized domain walls can account for the observed DM. 
We can express the amplitude $\phi_m$ in terms of the energy density associated with a domain-wall network $\rho_\textrm{TDN}$, the typical speed of a wall $v_\textrm{TD}$ (locally, $v_\textrm{TD} \sim 300~\textrm{km/s}$), and the average time between encounters of a system (e.g., Earth) with a wall $\mathcal{T}$: 
\begin{equation}
\label{TDN_energy_density}
\phi_m^2 \sim \rho_\textrm{TDN} v_\textrm{TD} \mathcal{T} d \, . 
\end{equation}
If domain walls of a single type comprise the entirety of the observed DM, then their largest dimension(s) cannot exceed the DM halo size of the smallest dwarf galaxies. 
This places the following upper bound on their transverse size:~$d \ll 1~\textrm{kpc}$, which can be relaxed if these objects make up only a sub-dominant fraction of the observed DM. 
In this section, we focus on the quadratic interactions of the field $\phi$ in (\ref{quad_portal}). 
We mention that one may also separately consider the case of Yukawa-type interactions of DM with SM matter and the effects of the resulting Yukawa force between a passing topological defect and the test masses of a detector, see \cite{Hall_GW-DM_2018} for more details.

\subsection{Experiments}
\label{Sec:TDM_Exp}
The passage of a domain wall through a laser-interferometric detector can result in similar signatures to those produced by conventional gravitational waves. 
One key difference between the two types of signatures is that domain walls are expected to pass through a detector with a relative speed of $v_\textrm{TD} \sim 10^{-3}$, rather than at the speed of light. 
Also, since the velocity distribution of domain walls is expected to be Maxwell-Boltzmannian (with a local average velocity of approximately zero), the ``event'' rate should be maximal for domain walls coming from the direction of the dark-matter ``wind'' (that is, from the direction towards which the Solar System is moving). 

The form of the output signal due to the passage of a domain wall through a laser interferometer will depend on several factors, including the size and geometry of the detector, the relative speed and direction of motion of the domain wall with respect to the detector, as well as the transverse size and cross-sectional profile of the wall. 
Rather than performing detailed numerical simulations of expected output signals, which we defer for future studies, let us consider the general features of domain-wall searches with laser interferometers, in order to estimate the sensitivities of these types of detectors to the quadratic interactions of the field $\phi$ in (\ref{quad_portal}). 

From Eqs.~(\ref{quad_VFCs}) and (\ref{TDN_energy_density}), we see that the magnitude of the domain-wall effects on the freely-suspended components of a ground-based detector (or freely-floating test masses of a space-based detector) scale as $\propto d$. 
For $d \ll L$, the output signal will generally contain appreciable power at and above the characteristic frequency of $f \sim v_\textrm{TD}/L$, which typically lies in the optimal frequency range of the detector. 
Additionally, there is generally no common-mode suppression in this case, in contrast to searches for oscillating DM fields (see Sec.~\ref{Sec:ODM}). 
On the other hand, for $d \gg L$, the output signal will generally be peaked at the characteristic frequency of $f \sim v_\textrm{TD}/d$, which typically lies well below the optimal frequency range of the detector, and there will also be common-mode suppression in this case. 
Hence we expect the sensitivity of a particular detector to the quadratic interactions of $\phi$ to be maximal for defects of transverse size $d \sim L$. 
We also point out that, in contrast to searches for oscillating DM fields, there is generally no $1/N_\textrm{eff}$ suppression of the output signal due to transient $\alpha$ or $m_e$ variations induced by a passing domain wall on a Fabry-Perot-Michelson interferometer (compare with Sec.~\ref{Sec:ODM_FP_exps}). 

For domain walls of transverse size greater than the thicknesses of the freely-suspended (or freely-floating) components of a detector, the wall can envelop the entirety of the regions of the components that are relevant for the interferometry measurement. 
In this case, from Eqs.~(\ref{quad_VFCs}), (\ref{size_changes1}) and (\ref{TDN_energy_density}), the maximum size change of a test mass is given by: 
\begin{equation}
\label{size_changes_TDM-max}
\left(\frac{\delta L_\textrm{test}}{L_\textrm{test}}\right)_\textrm{max} \sim - \rho_\textrm{TDN} v_\textrm{TD} \mathcal{T} d \left[ \left( \frac{1}{\Lambda'_\gamma} \right)^2 + \left( \frac{1}{\Lambda'_e} \right)^2 \right] \, . 
\end{equation}
For domain walls of transverse size less than the thicknesses of the freely-suspended (or freely-floating) components of a detector, the wall cannot envelop the components in their entirety. 
For a domain wall travelling parallel (in terms of its normal vector) to one of the arms of a ground-based detector, the size changes of the test masses in that arm will be suppressed compared to that in (\ref{size_changes_TDM-max}) by the factor $\sim d/L_\textrm{test} \ll 1$. 
On the other hand, for a domain wall travelling perpendicular to one of the arms of a ground-based detector, the size changes of that arm's test masses in the region through which the laser beam traverses will only be suppressed if the transverse size of the domain wall is smaller than the diameter of the laser beam. 
In the latter case, the duration of the passage will also be smaller compared with the case when the domain wall travels parallel to the arm. 

The freely-suspended components of a ground-based detector (or freely-floating test masses of a space-based detector) will also experience transient accelerations in accordance with Eq.~(\ref{test_mass_acceleration_general}). 
The passage of a domain wall with $d \gg L_\textrm{test}$ through a test mass occurs within a time interval of $\Delta t \sim d/v_\textrm{TD}$. 
In this case, the freely-suspended silica components of a ground-based detector will undergo the following maximum centre-of-mass displacements: 
\begin{align}
\label{test_mass_displacement_ground-silica_TDM}
\left| \delta \v{x}_\textrm{test} \right|_\textrm{max} &\sim \frac{\rho_\textrm{TDN} \mathcal{T} d^2}{v_\textrm{TD}}   \notag \\
&\times \left[ \frac{1.4 \times 10^{-3}}{\left(\Lambda'_\gamma\right)^2} + \frac{3 \times 10^{-4}}{\left(\Lambda'_e\right)^2} + \frac{1}{\left(\Lambda'_N\right)^2} \right]  \, , 
\end{align}
while on the other hand the Au-Pt test cubes of LISA will undergo the following maximum centre-of-mass displacements: 
\begin{align}
\label{test_mass_displacement_LISA_TDM}
\left| \delta \v{x}_\textrm{test} \right|_\textrm{max} &\sim \frac{\rho_\textrm{TDN} \mathcal{T} d^2}{v_\textrm{TD}}   \notag \\
&\times \left[ \frac{4 \times 10^{-3}}{\left(\Lambda'_\gamma\right)^2} + \frac{2 \times 10^{-4}}{\left(\Lambda'_e\right)^2} + \frac{1}{\left(\Lambda'_N\right)^2} \right]  \, , 
\end{align}
where we have again used Eqs.~(\ref{quad_VFCs}) and (\ref{TDN_energy_density}). 
On the other hand, the passage of a domain wall with $d \ll L_\textrm{test}$ through a test mass occurs within a time interval of $\Delta t \sim L_\textrm{test}/v_\textrm{TD}$, but the test mass (more precisely, a small portion $\sim d/L_\textrm{test} \ll 1$ thereof) sees an appreciably non-zero gradient of the domain wall only during a total time interval of $\Delta t \sim d/v_\textrm{TD}$. 
In this case, the maximum centre-of-mass displacement of a test mass will be suppressed compared to that in (\ref{test_mass_displacement_ground-silica_TDM}) or (\ref{test_mass_displacement_LISA_TDM}) by the factor $\sim d/L_\textrm{test} \ll 1$. 

\begin{figure*}[h!]
\centering
\includegraphics[height=0.35\linewidth]{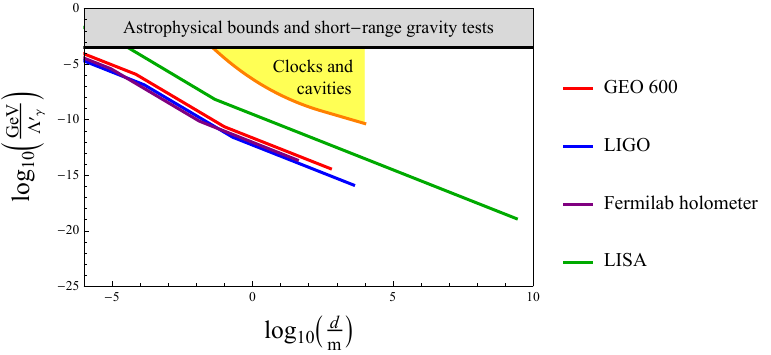}
\par\bigskip
\includegraphics[height=0.35\linewidth]{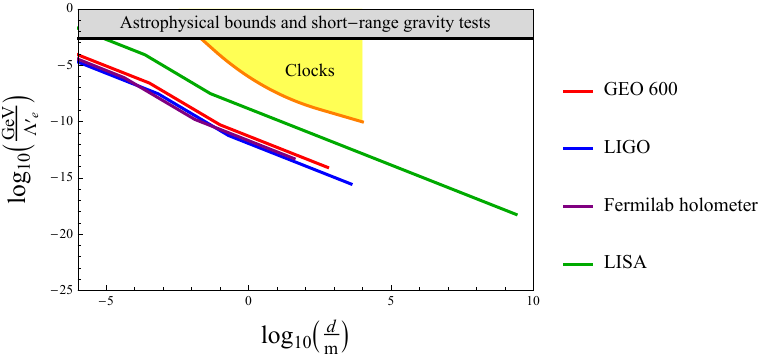}
\par\bigskip
\includegraphics[height=0.35\linewidth]{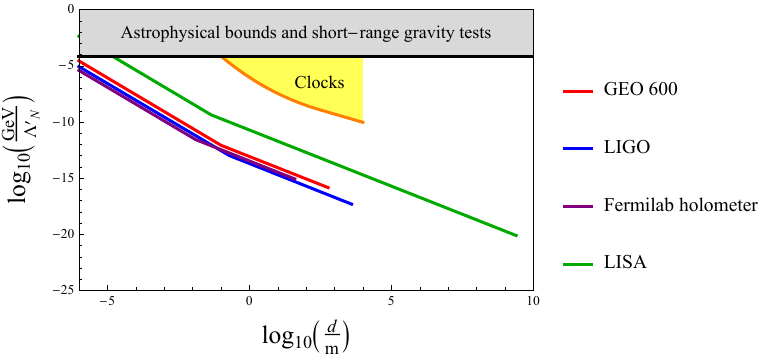}
\caption{(Color online). 
From top to bottom:~Physical parameter spaces for the quadratic interactions of a domain-wall field $\phi$ with the electromagnetic field (photon), electron and nucleons, as functions of the transverse size of a domain wall $d$. 
The solid lines denote the estimated sensitivities of current ground-based laser interferometers (red = GEO\,600, blue = LIGO, purple = Fermilab holometer). 
The solid green line denotes the projected sensitivity of the space-based LISA interferometer. 
All of these sensitivities assume a domain-wall transit speed of $v_\textrm{TD} \sim 300~\textrm{km/s}$, an average time between encounters of Earth and a domain wall of $\mathcal{T} \sim 1~\textrm{year}$, and that the domain-wall network saturates the average local cold DM density of $\rho_\textrm{DM} \approx 0.4~\textrm{GeV}/\textrm{cm}^3$. 
The region in grey denotes existing model-independent constraints from astrophysical observations \cite{Pospelov_2008_gen} and short-range tests of gravity \cite{Pospelov_2008_gen,Adelberger_2007_r3}. 
The regions in yellow denote existing constraints from domain-wall searches using a pair of co-located clocks referenced to a common cavity \cite{Wcislo_2016_cavity-DM} and via networks of clocks \cite{Roberts_2017_GPS-DM,Wcislo_2018_TDM}. 
}
\label{Fig:TDM_sensitivities}
\end{figure*}

Using Eqs.~(\ref{quad_VFCs}), (\ref{size_changes_TDM-max}), (\ref{test_mass_displacement_ground-silica_TDM}) and (\ref{test_mass_displacement_LISA_TDM}), and noting the suppression factors for thin domain walls discussed above, we estimate the current sensitivities of GEO\,600 \cite{GEO600_2013,GEO600_2015}, LIGO \cite{aLIGO_2016} and the Fermilab holometer using both of its co-located interferometers \cite{Fermilab_holometer_2016,Fermilab_holometer_2017}, as well as the projected sensitivity of LISA \cite{LISA_2017}, to the quadratic interactions of the domain-wall field $\phi$ with the photon, electron and nucleons in (\ref{quad_portal}). 
We present these estimates as solid lines in Fig.~\ref{Fig:TDM_sensitivities} (red = GEO\,600, blue = LIGO, purple = Fermilab holometer using both of its co-located interferometers, green = LISA), assuming an average time between encounters of Earth and a domain wall of $\mathcal{T} \sim 1~\textrm{year}$ and that the domain-wall network saturates the average local cold DM density of $\rho_\textrm{DM} \approx 0.4~\textrm{GeV}/\textrm{cm}^3$. 
For simplicity, in arriving at these estimates, we have neglected any possible excitation of underlying mechanical resonances associated with the detector that may arise due to the passage of a sufficiently thin (and/or quickly moving) domain wall. 
Furthermore, in the case of the Fermilab holometer, we have also neglected the fact that when the transverse size of a passing domain wall is smaller than the spatial separation between the two independent interferometers ($0.9~\textrm{m}$ in the current setup), if the domain wall at some moment in time simultaneously envelops one pair of test masses, then at another moment in time the domain wall might not simultaneously envelop another pair of test masses. 
In this case, the cross-correlation signal would be diminished; however, the individual interferometers would still respond in the usual manner described above. 
We note that, in contrast to oscillating DM fields (see Sec.~\ref{Sec:ODM}), the transient centre-of-mass displacements of the test masses due to the passage of a domain wall generally dominate over the transient size changes of the test masses, not only in space-based detectors but also in ground-based detectors, since in the case of domain walls of transverse size $d \ll L$ there is generally no common-mode suppression and hence no suppression factor of the form $\sim L/\lambdabar_\textrm{dB} \ll 1$.

\section{Discussion}
\label{Sec:Discussion}
In this paper, we have pointed out and explored in detail several new effects of DM on the components of laser interferometers. 
The estimated sensitivities of existing, modified and future laser-interferometry experiments to oscillating DM fields and domain walls are presented in Figs.~\ref{Fig:ODM_sensitivities} and \ref{Fig:TDM_sensitivities}, respectively. 
We see that existing ground-based laser interferometers already have sufficient sensitivity to probe extensive regions of unconstrained parameter space in both of these DM models.

In the case of oscillating DM fields, Michelson interferometers are especially sensitive. 
In particular, the existing GEO\,600 detector already offers up to 2 orders of magnitude more sensitivity than the best current (non-DM-based) constraints from fifth-force experiments \cite{MICROSCOPE_2017,MICROSCOPE_2018,Torsion_balance_Uranium_1999,Stadnik_2016_5F} in the scalar particle mass range $\textrm{few} \times 10^{-13}~\textrm{eV} \lesssim m_\phi \lesssim \textrm{few} \times 10^{-11}~\textrm{eV}$, and at least 8 orders of magnitude more sensitivity than recent DM searches via clock-cavity comparison experiments \cite{Aharony_2019_cavity-DM} in the same mass range. 
For scalar particle masses in the range $\textrm{several} \times 10^{-11}~\textrm{eV} \lesssim m_\phi \lesssim \textrm{several} \times 10^{-9}~\textrm{eV}$, dedicated resonant narrowband searches using small-scale Michelson interferometers operating near room temperature may improve in sensitivity by up to 2 orders of magnitude compared with previous fifth-force experiments and by at least 6 orders of magnitude compared with the recent DM searches of Refs.~\cite{Aharony_2019_cavity-DM,Savalle_2019_cavity-DM,Tretiak_2019_cavity-DM}. 
The sensitivity of Fabry-Perot-Michelson interferometers to oscillating DM fields can be increased by making the thicknesses of the freely-suspended Fabry-Perot arm mirrors different in the two arms of the interferometer, offering up to 5 orders of magnitude more sensitivity than previous fifth-force experiments for a relative difference in thickness of only $10\%$ and using a pair of LIGO interferometers operating at the design sensitivity of Advanced LIGO. 
The sensitivity of LISA to oscillating DM fields can be increased by replacing some of the freely-floating Au-Pt alloy test masses by test masses made of much lighter elements, offering up to a few-dozen times more sensitivity than previous fifth-force experiments.

In the case of domain walls, existing ground-based laser interferometers are particularly sensitive to domain walls with transverse sizes of up to several km, offering a sensitivity of up to 8 orders of magnitude beyond all other existing experiments and measurements \cite{Wcislo_2016_cavity-DM,Pospelov_2008_gen,Adelberger_2007_r3,Roberts_2017_GPS-DM,Wcislo_2018_TDM}. 
The Fermilab holometer, with its two co-located Michelson interferometers, and especially a global network of laser interferometers would benefit from their ability to disentangle correlated domain-wall-induced signatures from uncorrelated noise sources. 
The space-based LISA interferometer, with its enormous ``aperture size'', will be sensitive to domain walls with transverse sizes of up to several million km. 

We emphasise that our newly suggested signatures of DM in laser interferometers scale to the first power of the underlying DM interaction parameters ($\propto 1/\Lambda_X$ in the case of linear interactions and $\propto 1/(\Lambda'_X)^2$ in the case of quadratic interactions), whereas conventional non-DM signatures scale to the second power of the same interaction parameters ($\propto 1/\Lambda_X^2$ in the case of linear interactions and $\propto 1/(\Lambda'_X)^4$ in the case of quadratic interactions). 
This more favourable scaling for our proposed searches will be especially advantageous for improving the sensitivities of future laser interferometers to similar DM signatures. 

Finally, we briefly discuss the issue of technical naturalness, which formally requires the corrections to the scalar particle mass $m_\phi$ from radiative processes involving the non-gravitational interactions in (\ref{lin_portal}) and (\ref{quad_portal}) to be smaller than the ``bare'' mass contribution. 
For the linear couplings in (\ref{lin_portal}), the 1-loop corrections to $m_\phi^2$ are $\delta m_\phi^2 \sim [m_e \Lambda / (4 \pi \Lambda_e)]^2$ for the scalar-electron coupling and $\delta m_\phi^2 \sim [\Lambda^2 / (4 \pi \Lambda_\gamma)]^2$ for the scalar-photon coupling, where $\Lambda$ is a new-physics cut-off scale, which we assume to be independent of the other mass-energy scales appearing in (\ref{lin_portal}). 
Technical naturalness thereby requires $\Lambda_e \gtrsim m_e \Lambda / (4\pi m_\phi)$ for the scalar-electron coupling and $\Lambda_\gamma \gtrsim \Lambda^2 / (4\pi m_\phi)$ for the scalar-photon coupling. 
We present these technically-natural regions as pale green regions in Fig.~\ref{Fig:ODM_sensitivities} for $\Lambda \sim 10~\textrm{TeV}$. 
We see that existing and modified ground-based laser interferometers have the ability to probe sizeable regions of technically-natural parameter space for the scalar-electron coupling.

\section*{Acknowledgments}
We are grateful to Valery Frolov for helpful discussions on the reference cavities of laser-interferometric gravitational-wave detectors and the signal-to-noise ratio scaling of resonant narrowband experiments, as well as helpful comments on the manuscript. 
We thank Harald L\"uck for comments on the cross-correlation analysis of co-located interferometers. 
We thank Kenneth A.~Strain and GariLynn Billingsley for information about the LIGO test masses. 
We thank Peter Wolf for helpful discussions. 
Y.V.S.~was supported by the Humboldt Research Fellowship from the Alexander von Humboldt Foundation.



\begin{thebibliography}{99}

\bibitem{PDG_2018_review} M.~Tanabashi \textit{et al}.~(Particle Data Group), \textit{Review of Particle Physics}, Phys.~Rev.~D \textbf{98}, 030001 (2018). 

\bibitem{Stadnik_2018_review} Y.~V.~Stadnik and V.~V.~Flambaum, \textit{Searches for New Particles Including Dark Matter with Atomic, Molecular and Optical Systems}, arXiv:1806.03115. 

\bibitem{Safronova_2018_review} M.~S.~Safronova, D.~Budker, D.~DeMille, D.~F.~J. Kimball, A.~Derevianko and C.~W.~Clark, \textit{Search for new physics with atoms and molecules}, Rev.~Mod.~Phys.~\textbf{90}, 025008 (2018). 

\bibitem{Footnote1} We stress the term `apparent' here, since DM fields are external fields that perturb physical systems, in a similar manner to how, say, an external electric field perturbs the energy levels of an atom. 

\bibitem{Stadnik_2015_DM-LI} Y.~V.~Stadnik and V.~V.~Flambaum, \textit{Searching for Dark Matter and Variation of Fundamental Constants with Laser and Maser Interferometry}, Phys.~Rev.~Lett.~\textbf{114}, 161301 (2015). 

\bibitem{Stadnik_2015_DM-VFCs} Y.~V.~Stadnik and V.~V.~Flambaum, \textit{Can Dark Matter Induce Cosmological Evolution of the Fundamental Constants of Nature?}, Phys.~Rev.~Lett.~\textbf{115}, 201301 (2015). 

\bibitem{Derevianko_2014_TDM} A.~Derevianko and M.~Pospelov, \textit{Hunting for topological dark matter with atomic clocks}, Nature Phys.~\textbf{10}, 933 (2014). 

\bibitem{Stadnik_2014_TDM} Y.~V.~Stadnik and V.~V.~Flambaum, \textit{Searching for Topological Defect Dark Matter via Nongravitational Signatures}, Phys.~Rev.~Lett.~\textbf{113}, 151301 (2014). 

\bibitem{Stadnik_2016_cavities} Y.~V.~Stadnik and V.~V.~Flambaum, \textit{Enhanced effects of variation of the fundamental constants in laser interferometers and application to dark-matter detection}, Phys.~Rev.~A \textbf{93}, 063630 (2016). 

\bibitem{Wcislo_2016_cavity-DM} P.~Wcislo, P.~Morzynski, M.~Bober, A.~Cygan, D.~Lisak, R.~Ciurylo and M.~Zawada, \textit{Experimental constraint on dark matter detection with optical atomic clocks}, Nature Astron.~\textbf{1}, 0009 (2016). 

\bibitem{Ye_2018_cavity-DM} C.~Kennedy, E.~Oelker, T.~Bothwell, D.~Kedar, L.~Sonderhouse, E.~Marti, S.~Bromley, J.~Robinson and J.~Ye, \textit{Constraints on Ultralight Dark Matter with an Optical Lattice Clock}, Bulletin of the American Physical Society, H06.00005 (2018). 

\bibitem{Aharony_2019_cavity-DM} S.~Aharony, N.~Akerman, R.~Ozeri, G.~Perez, I.~Savoray and R.~Shaniv, \textit{Constraining Rapidly Oscillating Scalar Dark Matter Using Dynamic Decoupling}, arXiv:1902.02788. 


\bibitem{GEO600_2013} H.~Grote, K.~Danzmann, K.~L.~Dooley, R.~Schnabel, J.~Slutsky and H.~Vahlbruch, \textit{First Long-Term Application of Squeezed States of Light in a Gravitational-Wave Observatory}, Phys.~Rev.~Lett.~\textbf{110}, 181101 (2013). 

\bibitem{GEO600_2015} K.~L.~Dooley \textit{et al}., \textit{GEO 600 and the GEO-HF upgrade program:~successes and challenges}, Class.~Quantum Grav.~\textbf{33}, 075009 (2016). 

\bibitem{Fermilab_holometer_2016} A.~S.~Chou \textit{et al}., \textit{First Measurements of High Frequency Cross-Spectra from a Pair of Large Michelson Interferometers}, Phys.~Rev.~Lett.~\textbf{117}, 111102 (2016). 

\bibitem{Fermilab_holometer_2017} A.~S.~Chou \textit{et al}., \textit{The Holometer:~an instrument to probe Planckian quantum geometry}, Class.~Quantum Grav.~\textbf{34}, 065005 (2017). 

\bibitem{aLIGO_2016} D.~V.~Martynov \textit{et al}., \textit{Sensitivity of the Advanced LIGO detectors at the beginning of gravitational wave astronomy}, Phys.~Rev.~D \textbf{93}, 112004 (2016); Erratum Phys.~Rev.~D \textbf{97}, 059901 (2018). 

\bibitem{aVirgo_2014} F.~Acernese \textit{et al}., \textit{Advanced Virgo:~a second-generation interferometric gravitational wave detector}, Class.~Quantum Grav.~\textbf{32}, 024001 (2015). 

\bibitem{KAGRA_2013} Y.~Aso, Y.~Michimura, K.~Somiya, M.~Ando, O.~Miyakawa, T.~Sekiguchi, D.~Tatsumi and H.~Yamamoto (The KAGRA Collaboration), \textit{Interferometer design of the KAGRA gravitational wave detector}, Phys.~Rev.~D \textbf{88}, 043007 (2013). 


\bibitem{LIGO_discovery_2016} B.~P.~Abbott \textit{et al}.~(LIGO Scientific Collaboration and Virgo Collaboration), \textit{Observation of Gravitational Waves from a Binary Black Hole Merger}, Phys.~Rev.~Lett.~\textbf{116}, 061102 (2016). 

\bibitem{LIGO_discovery_2017} B.~P.~Abbott \textit{et al}.~(LIGO Scientific Collaboration and Virgo Collaboration), \textit{GW170817:~Observation of Gravitational Waves from a Binary Neutron Star Inspiral}, Phys.~Rev.~Lett.~\textbf{119}, 161101 (2017). 

\bibitem{LISA_2017} K.~Danzmann \textit{et al}., \textit{LISA:~Laser Interferometer Space Antenna. A proposal in response to the ESA call for L3 mission concepts}, (2017). 


\bibitem{Piazza_2010_Higgs} F.~Piazza and M.~Pospelov, \textit{Sub-eV scalar dark matter through the super-renormalizable Higgs portal}, Phys.~Rev.~D \textbf{82}, 043533 (2010). 

\bibitem{Stadnik_2016_Higgs} Y.~V.~Stadnik and V.~V.~Flambaum, \textit{Improved limits on interactions of low-mass spin-0 dark matter from atomic clock spectroscopy}, Phys.~Rev.~A \textbf{94}, 022111 (2016). 


\bibitem{Flambaum_2018_sizes} L.~F.~Pasteka, Y.~Hao, A.~Borschevsky, V.~V.~Flambaum and P.~Schwerdtfeger, \textit{Material Size Dependence on Fundamental Constants}, Phys.~Rev.~Lett.~\textbf{122}, 160801 (2019). 

\bibitem{footnote_Sep2019} We note that an externally-driven damped harmonic oscillator and a parametrically-driven damped harmonic oscillator exhibit the same qualitative behaviour in the present context. 



\bibitem{Footnote0} The finite thicknesses of the power recycling mirror and end arm mirrors, which form the reference cavity for the laser, do in fact give very small corrections to $\delta \omega / \omega$ and $\delta n / n$. 
These small corrections to $\delta \omega / \omega$, coupled with the deliberately small geometric arm length difference in actual laser interferometers, also give a negligibly small contribution to the output signal. 


\bibitem{Footnote0B} We note that anisotropic inhomogeneities in the elemental composition of a test mass can also result in an acceleration on the test mass without there necessarily being a spatial gradient in $\alpha$ or the particle masses. 
However, for high-quality test masses that are made of practically homogeneous materials, the resulting observable effects are generally subleading compared to those due to spatial gradients in $\alpha$ or the particle masses. 


\bibitem{Vac_misal_A} J.~Preskill, M.~B.~Wise and F.~Wilczek, \textit{Cosmology of the invisible axion}, Phys.~Lett.~B \textbf{120}, 127 (1983). 

\bibitem{Vac_misal_B} L.~F.~Abbott and P.~Sikivie, \textit{A cosmological bound on the invisible axion}, Phys.~Lett.~B \textbf{120}, 133 (1983). 

\bibitem{Vac_misal_C} M.~Dine and W.~Fischler, \textit{The not-so-harmless axion}, Phys.~Lett.~B \textbf{120}, 137 (1983). 

\bibitem{Footnote2} In order for particles to form a classical field, there must be a large number of such particles within their reduced de Broglie volume; i.e., $n_\phi \lambdabar_\textrm{dB}^3 \gg 1$. 
For DM particles that saturate the observed average local cold DM density of $\rho_\textrm{DM} \approx 0.4~\textrm{GeV}/\textrm{cm}^3$, this requirement is readily satisfied by DM particles with sub-eV masses. 


\bibitem{Derevianko_DM-lineshape_2018} A.~Derevianko, \textit{Detecting dark-matter waves with a network of precision-measurement tools}, Phys.~Rev.~A \textbf{97}, 042506 (2018). 


\bibitem{Hees_2018_DM} A.~Hees, O.~Minazzoli, E.~Savalle, Y.~V.~Stadnik and P.~Wolf, \textit{Violation of the equivalence principle from light scalar dark matter}, Phys.~Rev.~D \textbf{98}, 064051 (2018). 


\bibitem{MICROSCOPE_2017} P.~Touboul, G.~Metris, M.~Rodrigues, Y.~Andre, Q.~Baghi, J.~Berge, D.~Boulanger, S.~Bremer \textit{et al}., \textit{MICROSCOPE Mission:~First Results of a Space Test of the Equivalence Principle}, Phys.~Rev.~Lett.~\textbf{119}, 231101 (2017). 

\bibitem{MICROSCOPE_2018} J.~Berge, P.~Brax, G.~Metris, M.~Pernot-Borras, P.~Touboul and J.-P.~Uzan, \textit{MICROSCOPE Mission:~First Constraints on the Violation of the Weak Equivalence Principle by a Light Scalar Dilaton}, Phys.~Rev.~Lett.~\textbf{120}, 141101 (2018). 

\bibitem{Torsion_balance_Uranium_1999} G.~L.~Smith, C.~D.~Hoyle, J.~H.~Gundlach, E.~G.~Adelberger, B.~R.~Heckel and H.~E.~Swanson, \textit{Short-range tests of the equivalence principle}, Phys.~Rev.~D \textbf{61}, 022001 (1999). 

\bibitem{Stadnik_2016_5F} N.~Leefer, A.~Gerhardus, D.~Budker, V.~V.~Flambaum and Y.~V.~Stadnik, \textit{Search for the Effect of Massive Bodies on Atomic Spectra and Constraints on Yukawa-Type Interactions of Scalar Particles}, Phys.~Rev.~Lett.~\textbf{117}, 271601 (2016). 

\bibitem{Leefer_2015_Dy-DM} K.~Van Tilburg, N.~Leefer, L.~Bougas and D.~Budker, \textit{Search for Ultralight Scalar Dark Matter with Atomic Spectroscopy}, Phys.~Rev.~Lett.~\textbf{115}, 011802 (2015). 

\bibitem{Hees_2016_clock-DM} A.~Hees, J.~Guena, M.~Abgrall, S.~Bize and P.~Wolf, \textit{Searching for an Oscillating Massive Scalar Field as a Dark Matter Candidate Using Atomic Hyperfine Frequency Comparisons}, Phys.~Rev.~Lett.~\textbf{117}, 061301 (2016). 

\bibitem{Savalle_2019_cavity-DM} E.~Savalle, B.~M.~Roberts, F.~Frank, P.-E.~Pottie, B.~T.~McAllister, C.~B.~Dailey, A.~Derevianko and P.~Wolf, \textit{Novel approaches to dark-matter detection using space-time separated clocks}, arXiv:1902.07192. 

\bibitem{Tretiak_2019_cavity-DM} D.~Antypas, O.~Tretiak, A.~Garcon, R.~Ozeri, G.~Perez and D.~Budker, \textit{Scalar dark matter in the radio-frequency band:~atomic-spectroscopy search results}, arXiv:1905.02968. 

\bibitem{AURIGA_2017_DM} A.~Branca \textit{et al}., \textit{Search for an Ultralight Scalar Dark Matter Candidate with the AURIGA Detector}, Phys.~Rev.~Lett.~\textbf{118}, 021302 (2017). 


\bibitem{GW-Dark_photon_2019} H.-K.~Guo, K.~Riles, F.-W.~Yang and Y.~Zhao, \textit{Searching for Dark Photon Dark Matter in LIGO O1 Data}, arXiv:1905.04316. 


\bibitem{Footnote_LIGO_CC} The spatial separation between the two LIGO detectors is $3000~\textrm{km}$, and so an oscillating DM field is spatially coherent over both detectors for all DM particle masses of interest. 


\bibitem{Arvanitaki2016_resonance-bar} A.~Arvanitaki, S.~Dimopoulos, and K.~Van Tilburg, \textit{Sound of Dark Matter:~Searching for Light Scalars with Resonant-Mass Detectors}, Phys.~Rev.~Lett.~\textbf{116}, 031102 (2016). 


\bibitem{Footnote4} We note that for phenomenologically interesting interaction strengths of an oscillating DM field with a quality factor of $Q_\phi \sim 10^6$, even resonantly-enhanced DM-induced cavity length fluctuations are not sufficiently large to change the standing-wave mode number of the laser field inside the cavity (typical mode numbers are $\sim 10^6$). 

\bibitem{Footnote4b} Obviously, `exchanging' test masses on a satellite mission such as LISA would be the scope of a new mission. 

\bibitem{Footnote5} We note that, for the optimal frequency ranges of ground-based laser interferometers, the non-common-mode components of DM-induced time-varying displacements of the parts of Earth directly in contact with the structures connected to the pivot points of two different test-mass systems correspond to time-varying size changes of (the solid) Earth at frequencies well above the fundamental vibrational frequency of Earth, and so are strongly suppressed (as are isotropic DM-induced time-varying size changes of Earth), meaning that the pivot points are practically unaffected by the DM interactions under consideration. 
Furthermore, the optimal frequency ranges of ground-based detectors lie well above the normal-mode frequencies of their double-pendula suspension systems for the test masses, meaning that the test masses respond to DM-induced time-varying forces as though they were free (i.e., the test masses respond inertially in this case). 


\bibitem{GW-Dark_photon_2018} A.~Pierce, K.~Riles and Y.~Zhao, \textit{Searching for Dark Photon Dark Matter with Gravitational-Wave Detectors}, Phys.~Rev.~Lett.~\textbf{121}, 061102 (2018). 


\bibitem{Arvanitaki2015_dilaton-GW} A.~Arvanitaki, J.~Huang, and K.~Van Tilburg, \textit{Searching for dilaton dark matter with atomic clocks}, Phys.~Rev.~D \textbf{91}, 015015 (2015). 

\bibitem{Suyama2015_dilaton-GW} S.~Morisaki and T.~Suyama, \textit{On the detectability of ultralight scalar field dark matter with gravitational-wave detectors}, arXiv:1811.05003. 


\bibitem{nEDM_2017_axion-DM} C.~Abel, N.~J.~Ayres, G.~Ban, G.~Bison, K.~Bodek, V.~Bondar, M.~Daum, M.~Fairbairn, V.~V.~Flambaum \textit{et al}., \textit{Search for Axionlike Dark Matter through Nuclear Spin Precession in Electric and Magnetic Fields}, Phys.~Rev.~X \textbf{7}, 041034 (2017). 


\bibitem{Pitjeva_2013_DM-overdensity} E.~V.~Pitjeva and N.~P.~Pitjev, \textit{Relativistic effects and dark matter in the Solar system from observations of planets and spacecraft}, MNRAS \textbf{432}, 3431 (2013). 

\bibitem{Adler_2008_DM-overdensity} S.~L.~Adler, \textit{Placing direct limits on the mass of earth-bound dark matter}, J.~Phys.~A \textbf{41}, 412002 (2008). 


\bibitem{Eby_DM-star_2019} A.~Banerjee, D.~Budker, J.~Eby, H.~Kim and G.~Perez, \textit{Relaxion Stars and their detection via Atomic Physics}, arXiv:1902.08212. 


\bibitem{Vilenkin_1985_TDs} A.~Vilenkin, \textit{Cosmic strings and domain walls}, Phys.~Rep.~\textbf{121}, 263 (1985). 


\bibitem{Hall_GW-DM_2018} E.~D.~Hall, R.~X.~Adhikari, V.~V.~Frolov, H.~Mueller and M.~Pospelov, \textit{Laser interferometers as dark matter detectors}, Phys.~Rev.~D \textbf{98}, 083019 (2018). 


\bibitem{Pospelov_2008_gen} K.~A.~Olive and M.~Pospelov, \textit{Environmental dependence of masses and coupling constants}, Phys.~Rev.~D \textbf{77}, 043524 (2008). 

\bibitem{Adelberger_2007_r3} E.~G.~Adelberger, B.~R.~Heckel, S.~Hoedl, C.~D.~Hoyle, D.~J.~Kapner and A.~Upadhye, \textit{Particle-Physics Implications of a Recent Test of the Gravitational Inverse-Square Law}, Phys.~Rev.~Lett.~\textbf{98}, 131104 (2007). 

\bibitem{Roberts_2017_GPS-DM} B.~M.~Roberts, G.~Blewitt, C.~Dailey, M.~Murphy, M.~Pospelov, A.~Rollings, J.~Sherman, W.~Williams and A.~Derevianko, \textit{Search for domain wall dark matter with atomic clocks on board global positioning system satellites}, Nature Commun.~\textbf{8}, 1195 (2017). 

\bibitem{Wcislo_2018_TDM} P.~Wcislo \textit{et al}., \textit{New bounds on dark matter coupling from a global network of optical atomic clocks}, Sci.~Adv.~\textbf{4}, eaau4869 (2018). 




\end{thebibliography}

\end{document}